# A second planetesimal collision in the Fomalhaut system

Paul Kalas[1,2,3]\*, Jason J. Wang[4,5], Maxwell A. Millar-Blanchaer[6], Bin B. Ren[7,8,9]\*, Mark C. Wyatt[10], Grant M. Kennedy[11,12], Maximilian Sommer[10], Thomas M. Esposito[1,2], Robert J. De Rosa[13], Michael Fitzgerald[14]

[1]Astronomy Department, University of California, Berkeley, CA, USA. [2]Search for Extraterrestrial Intelligence Institute, Carl Sagan Center, Mountain View, CA, USA. [3]Institute of Astrophysics, Foundation for Research and Technology Hellas, Heraklion, Greece. [4]Center for Interdisciplinary Exploration and Research in Astrophysics, Northwestern University, Evanston, IL, USA. [5]Department of Physics and Astronomy, Northwestern University, Evanston, IL, USA. [6]Department of Physics, University of California, Santa Barbara, CA, USA. [7]Department of Astronomy, Xiamen University, Xiamen, China. [8]Observatoire de la Côte d'Azur, Université Côte d'Azur, Nice, France. [9]Max-Planck-Institut für Astronomie, Heidelberg, Germany. [10]Institute of Astronomy, University of Cambridge, Cambridge, UK. [11]Malaghan Institute of Medical Research, Wellington, New Zealand. [12]Department of Physics, University of Warwick, Coventry, UK. [13]European Southern Observatory, Santiago, Chile. [14]Department of Physics and Astronomy, University of California, Los Angeles, CA, USA.

\*Corresponding author. Email: kalas@berkeley.edu (P.K.); rbb@xmu.edu.cn (B.B.R.)

**The nearby star Fomalhaut is orbited by a compact source, Fomalhaut b, which has previously been interpreted as either a dust-enshrouded exoplanet or a dust cloud generated by the collision of two planetesimals. Such collisions are rarely observed but their debris can appear in direct imaging. We report Hubble Space Telescope observations that show the appearance in 2023 of a second point source around Fomalhaut, resembling the appearance of Fomalhaut b twenty years earlier. We interpret this additional source as a dust cloud produced by a recent impact between two planetesimals. The positions and motion of two impact-generated dust clouds over twenty years provide constraints on the collisional dynamics in the debris belt.**

The nearby star Fomalhaut (also known as HD 216956) is at a distance of 7.7 parsecs (pc) from the Sun, and has an age of 440 Myr (*1*). This star is orbited by an outer dusty debris belt (Fig. 1), which has a geometric center offset by ~15 astronomical units (au) from the star and a sharp inner edge, previously attributed to gravitational perturbations by an undetected planet (*2–4*). Coronagraphic optical observations with the Hubble Space Telescope (HST) in 2004 and 2006 showed a point source, designated Fomalhaut b, which had the same sky motion as the star (*5*). It appeared to orbit interior to the dust belt boundary of 133 au and exhibited possible optical variability. Fomalhaut b was brighter than predicted by thermal emission models of extrasolar planets, but was not detected in deep near-infrared observations (*5*).

Fomalhaut b has been interpreted as either an exoplanet surrounded by dust that reflects starlight, or an expanding dust cloud produced by a planetesimal collision (*6–8*). The designation Fomalhaut b would be appropriate for an extrasolar planet; however because the nature of the source is debated, we hereafter adopt the nomenclature Fomalhaut cs1 (Fom cs1), an abbreviation of 'circumstellar source 1'. If cs1 is a dust cloud, we expect radiation pressure and orbital shearing to disperse it over a timescale of years, decreasing cs1's surface brightness and increasing its spatial extent. However, measuring changes in the brightness and morphology of cs1 is difficult because it is faint [apparent visual magnitude $m_v \approx 25$ magnitude (mag)] and located near a much brighter star ($m_v = 1.2$ mag). Additional observations in 2010 and 2012 (Fig. 1A) showed that the orbit of cs1 is highly eccentric (*9*). Further observations in 2013 and 2014 appeared to show that it had expanded in size and faded (*10*). The position of cs1 in 2013 indicated that it had undergone radial acceleration, and cs1 was not detected at all in 2014 (*10*). This observed behavior is consistent with a collision between two planetesimals that occurred before 2004 and produced an expanding gravitationally unbound dust cloud, which eventually faded below the sensitivity of the observations. However, the 2014 observations had less sensitivity than prior epochs, which might explain the non-detection of cs1 at that epoch.

### Observation of an additional point source

We performed a deep search for cs1 in 2023 using HST's Space Telescope Imaging Spectrograph (STIS) (*11*), in coronographic imaging mode at four orientation angles (table S1). We used an occulting wedge and angular differential imaging to subtract the light from the star (*12*), producing an image of the debris disc (Fig. 1B). Four independent data reduction methods were employed, with consistent results (*12*).

Fom cs1 is not detected in the HST observations made in September 2023, but a different point source had appeared, which we designate Fomalhaut cs2 (Fom cs2). This source is detected with a signal-to-noise ratio (SNR) of 7 to 9 in the four independent data reduction processes (fig. S1). It is situated at the inner edge of the debris belt, with angular separation 12.76 ± 0.02" from the star and position angle (PA) = 311.27 ± 0.05°, clockwise from north (Fig. 1B, fig. S1, and table S2). This is 1.26" southwest of where cs1 was observed ~20 years earlier (Fig. 2), which was at separation 12.58" and PA=316.93° (*5*). If both sources and the outer dust belt lie in the same plane, cs1 was 120.6 au from the star in 2004 and cs2 was at 135.3 au radius in 2023. The angular separation between the two sources in the belt plane is 8.1° and the distance between them is 23.4 au.

The inner edge of the dust belt has radius ~133 au, though previous observations have shown dust-scattered light inward of this boundary (*4*), and another intermediate dust belt at ~94 au has been detected in infrared imaging (*13*). Previous HST observations had the sensitivity to detect Fom cs2, but did not do so (*9*, *10*). This indicates that cs2 is neither a background object (fig. S2) nor a pre-existing source orbiting within the dust belt. Previous infrared imaging at 3 to 4 μm obtained in October 2022 (*14*) did not have the necessary sensitivity to detect sources as faint as cs2. In 2023, Fom cs2 is 0.3 mag (130%) brighter than cs1 was in 2012. It could have appeared at any point in the nine years since the last HST observation in September 2014.

We obtained follow-up HST imaging in September 2024, which however has substantially lower sensitivity than the 2023 observation (*12*). These data show a candidate source with SNR ~3 that is displaced 110 ± 17 milliarcseconds (mas) north of Fom cs2's position in 2023 (figs. S3 and S4 and table S3). Given the 357.7 days between observations, this offset corresponds to sky plane motion of 0.113" yr$^{-1}$. If this candidate source is cs2, its position indicates a highly eccentric orbit, with eccentricity $e \sim 0.8$ (fig. S5 and table S4). The radial direction with respect to the star is 48.7° clockwise from north, yet there is no measurable east-west motion between cs2 in 2023 and the candidate source in 2024, so the two are inconsistent with purely radial motion.

We do not find conclusive evidence of Fom cs1 in the 2023 observation (fig. S6). We reanalyzed the 2013 HST observations and confirmed the previous measurement (*10*) of cs1's radial acceleration between 2012 and 2013 epochs (figs. S9 to S13 and tables S5 and S6). In the belt plane, cs1 had a radial velocity component of 3.4 km s$^{-1}$ from 2010 to 2012, which increased to 11.7 km s$^{-1}$ over the following year. This acceleration (1.946 × 10$^{-7}$ km s$^{-2}$) is consistent with theoretical expectations for radiation pressure acting on sub-micron dust grains (*10*). If that acceleration was sustained for a decade, the cloud of cs1 dust grains would be situated ~50 au farther away from the star (in the belt plane) in 2023 than it was in 2013. If this has occurred, Fom cs1 would be receiving ~50% less light from the star, reducing its brightness to the limit of our sensitivity. We identify a candidate source at this approximate location but cannot validate it (fig. S8).

**Collisions in the Fomalhaut system**

The appearance of Fom cs2 supports the interpretation that cs1 was a dust cloud from a planetesimal collision, not reflected light from dust around an exoplanet. We interpret both cs1 and cs2 as evidence of a collisional process in the Fomalhaut planetary system (*15*). Similar collisional debris has been directly observed in the Solar System, such as the disruptions of asteroids P/2010 A2 in 2009 (*16*, *17*) and (596) Scheila in 2010 (*18*, *19*). In extrasolar planetary systems, there is indirect evidence of collisions, including peculiar transit light curves (*20*, *21*), variable spectroscopic and thermal emission properties (*22*, *23*), and unusual dust structures (*24*–*26*).

Fom cs1 and cs2 occurred over a ~20 year timescale. We use these events to estimate the collision rate, the mass of dust grains generated, and the sizes of the planetesimals involved. Assuming a dust cloud with a grain density of 3.5 g cm$^{-3}$ and radius size distribution extending from 0.1 μm to 1 mm, the optical brightness of cs1 requires a total cross-sectional area of ~5 × 10$^{22}$ cm$^2$ of dust, with total mass ~10$^{20}$ g (*10*). This dust mass is nine orders of magnitude greater than the asteroid disruption events observed in the Solar System (*16*–*19*). However, the brightness is dominated by grains up to ~3 μm in radius, which contain ~10% of that mass or 10$^{19}$ g. Grains this small are subsequently ejected from the planetary system by stellar radiation pressure (*27*). Taking ~20 years as the mean timescale between events in the Fomalhaut system, this implies that there have been 22 million cs1-like events in the 440 Myr since its formation, which would have depleted the belt by 2.2 × 10$^{26}$ g [0.04 Earth masses ($M_{earth}$)].

However, the cs1-like collisional events themselves are not the dominant mass loss mechanism from the belt, which contains at least 1.8 $M_{earth}$ in bodies up to 0.3 km in radius (*28*). Assuming a steady state, maintaining the dust mass observed in the belt requires the release of grains from < 0.3 km-sized planetesimals at a rate ~45 times higher than we calculated for cs1-like events. Bodies larger than 0.3 km could also be present in the belt, but would predominantly be a primordial population, not collisional fragments produced by catastrophic destruction of larger bodies. Collisions among this primordial population are likely to be the origin of cs1-like events, because the mass of unbound dust for cs1-like events (calculated above) requires the disruption of bodies >10 km in radius. A previous model invoked the catastrophic disruption of ~100 km radius bodies (*10*). While such collisions would produce the unbound dust seen as cs1-like events, they would also produce a population of bound fragments of mass ~10$^{22}$ g per event. Because this would replenish the belt at a higher rate (50 $M_{earth}$ in 440 Myr) than it is currently losing mass through steady-state collisional erosion, we suggest that cs1-like events are more likely to be erosive collisions or involve bodies smaller than 100 km.

We use the 20-year timescale between events to constrain the size, $R$, of parent bodies of cs1-like events. A previous model of the Fomalhaut debris belt (*28*) estimated there are ~3 × 10$^{13}$ bodies of radius 0.3 km collisionally maintaining the dust belt (calculated as 1.8 $M_{earth}$ divided by the mass of a 0.3 km body). The mean time between collisions for each of these objects must be 440 Myr [the age of the system (*1*)], so we expect ~10$^6$ collisions every 20 years. Assuming that primordial bodies larger than 0.3 km have a power law size distribution with index $\alpha$, the collision lifetime within that population is then 440 Myr × $(R / 0.3 \text{ km})^{\alpha-3}$ and the number of objects in each logarithmic bin at size $R$ is $3 \times 10^{13} \times (R / 0.3 \text{ km})^{1-\alpha}$. For collisions that occur once every 20 years, the most probable size is $R \sim 30 \text{ km} \times 10^{(7-2\alpha)/(\alpha-2)}$ in radius, which have total mass $1.8 \text{ M}_{earth} \times (R / 0.3 \text{ km})^{4-\alpha}$. Adopting $\alpha = 3.5$ (29), this implies that the parent bodies of cs1-like events have radius 30 km and a total mass of 18 $M_{earth}$. There is sufficient mass in such bodies to explain the brightness of cs1-like events, if each collision results in ~4% of the mass being released as dust grains <3 µm in size. This calculation assumes a primordial size distribution with $\alpha = 3.5$, which has not been constrained by observations. Alternative values $\alpha = 3$ or 4 would imply parent bodies of radius 300 km or 10 km, respectively. The former would require 1800 $M_{earth}$ of planetesimals, and the latter would require that 100% of the planetesimal's mass is released as <3 µm-sized dust following the collision, both of which we regard as implausible.

Our calculation that 4% of the mass of 30 km planetesimals is released as small grains is three orders of magnitude higher than expected for a fragment size distribution with $\alpha=3.5$ extending up to planetesimal size. However, these large planetesimals would be expected to form a deep dusty regolith from many lower-energy impacts that shatter the bodies and the fragments reaccumulate on the surface due to gravity, rather than being dispersed away from the parent body. The material strength of a 30 km planetesimal is ~3600 times lower than the catastrophic dispersal threshold (30), so it can be fragmented by an impactor 15 times smaller than required for catastrophic destruction. Because the collision rate scales with impactor size as $R^{1-\alpha}$ (31), for $\alpha = 3.5$ we expect a 30 km planetesimal to have undergone ~900 shattering events before a catastrophic impact. While this is a simplified estimate, because shattering impacts erode some fraction of a planetesimal's mass (32), it indicates that there are many times more shattering collisions than catastrophic collisions.

Fom cs1 appeared to follow a Keplerian orbit between 2004 and 2012, before accelerating radially (figs. S11 to S13). This could be because the dust cloud remained optically thick for at least 8 years, before becoming optically thin. An optically thick dust cloud would experience radiation pressure only on the star-facing hemisphere. In an optically thin dust cloud, all the dust experiences radiation pressure and accelerates outwards. This process could become a self-reinforcing runaway, as more acceleration makes the cloud disperse further, becoming increasingly optically thin. This phenomenon has been proposed to explain the infrared variability of other debris disk systems (23). Under this scenario, we estimate the size of the dust cloud when the runaway acceleration begins. This occurs when the optical depth $\tau=1$, which is when the dust cloud has expanded to a radius at which the area of the cloud as seen by the star is equal to the total cross-sectional area of dust within it ($5 \times 10^{22}$ cm$^2$). For a linear expansion rate, the radius of the cs1 dust cloud when it becomes optically thin would be ~0.008 au ($1.3 \times 10^6$ km). If this optically thick phase was ~8 years, this implies an expansion rate of ~5 m s$^{-1}$, which is 12% of the escape velocity of a 30 km-radius body. For comparison, the mean velocity of fragments that escaped from Theia's collision with proto-Earth (which produced the Moon) was ~47% of Earth's escape velocity (33). This scenario implies that during the optically thick phase the cloud would become brighter with time, as it intercepts more radiation. This was not observed, though it might not have been detectable due to the substantial photometric uncertainties.

The temporal and spatial proximity of Fom cs1 and cs2 (Fig. 2) implies that the collisions might not be random. In the belt plane, more than half of the belt edge is obscured by residual noise or field of view limitations; we estimate that the observations are sensitive to cs1/cs2-like sources over 162° of the belt (12). The probability of finding a second source randomly located within 8.1° on either side of cs1 is then (8.1° × 2) / 162° = 10%. Although we cannot rule out chance coincidence, we consider the dynamical processes that could spatially concentrate cs1/cs2 events. Infrared observations have shown a misaligned intermediate dust belt, ~20 au wide and centered at ~94 au (12). It is inclined by 7.4 to 22.9°, relative to the outer belt, and has a higher eccentricity ($e = 0.3$). This misalignment means that any planetesimals ejected from the intermediate belt could intersect the outer belt at a specific azimuth, causing collisions at similar locations. However, the intersection between the two belt planes is at PA ~347° (table S3), which is ~70° away from the positions of both cs1 and cs2, disfavoring this scenario. An alternative dynamical pathway could involve planetesimals trapped in mean-motion resonances with an exoplanet, producing a higher number density and collision rate in the cs1/cs2 region (34). Previous observations of mm-sized dust grains in the Fomalhaut system do not show an over-density in this region (35), though radiation pressure could cause the orbits of mm-sized dust to not be spatially confined by mean motion resonances with Earth-mass planets (36).

## ACKNOWLEDGMENTS

**Funding:** PK, MMB, JJW, and TME were supported by grant HST-GO-17139 from the Space Telescope Science Institute under NASA contract NAS5-26555. MCW and MS were supported by the United Kingdom Research and Innovation / Science and Technology Facilities Council grants UKRI1198 and ST/W000997/1, respectively. BBR received funding from the European Union's Horizon Europe research and innovation program under the Marie Skłodowska-Curie grant agreement No. 101103114; from the National Science Foundation of China; and from Xiamen University starting grant X2450230. **Author contributions:** PK, JJW, MMB and BBR reduced and analyzed the HST data. PK, JJW, MMB, and BBR independently employed the data reduction methods 1 to 4, respectively. JJW fitted the Keplerian orbits. TME, RJDR, and MF assisted in planning the observations and interpreting measurements. MCW, GMK and MS contributed theoretical interpretation and comparison with prior literature. All authors contributed to writing and revising the manuscript. **Competing interests:** T.E. is also affiliated with Unistellar, Marseille, France and SkyMapper Inc., San Francisco, CA, USA. **Data and materials availability**: The HST observations are available in the Mikulski Archive for Space Telescopes at [https://mast.stsci.edu/search/ui/#/hst](https://mast.stsci.edu/search/ui/#/hst), under proposal IDs 9862 (2004 August epoch), 10390 (2004 October), 10598 (2006), 11818 (2010), 12576 (2012), 13037 (2013), 13726 (2014), and 17139 (2023 and 2024). The pyKLIP software is available at [https://bitbucket.org/pyKLIP/pyklip/src/main/](https://bitbucket.org/pyKLIP/pyklip/src/main/). No physical materials were generated in this work.

## SUPPLEMENTARY MATERIALS

[science.org/doi/10.1126/science.adu6266](science.org/doi/10.1126/science.adu6266)
Materials and Methods
Figs. S1 to S13
Tables S1 to S6
References (*37–53*)

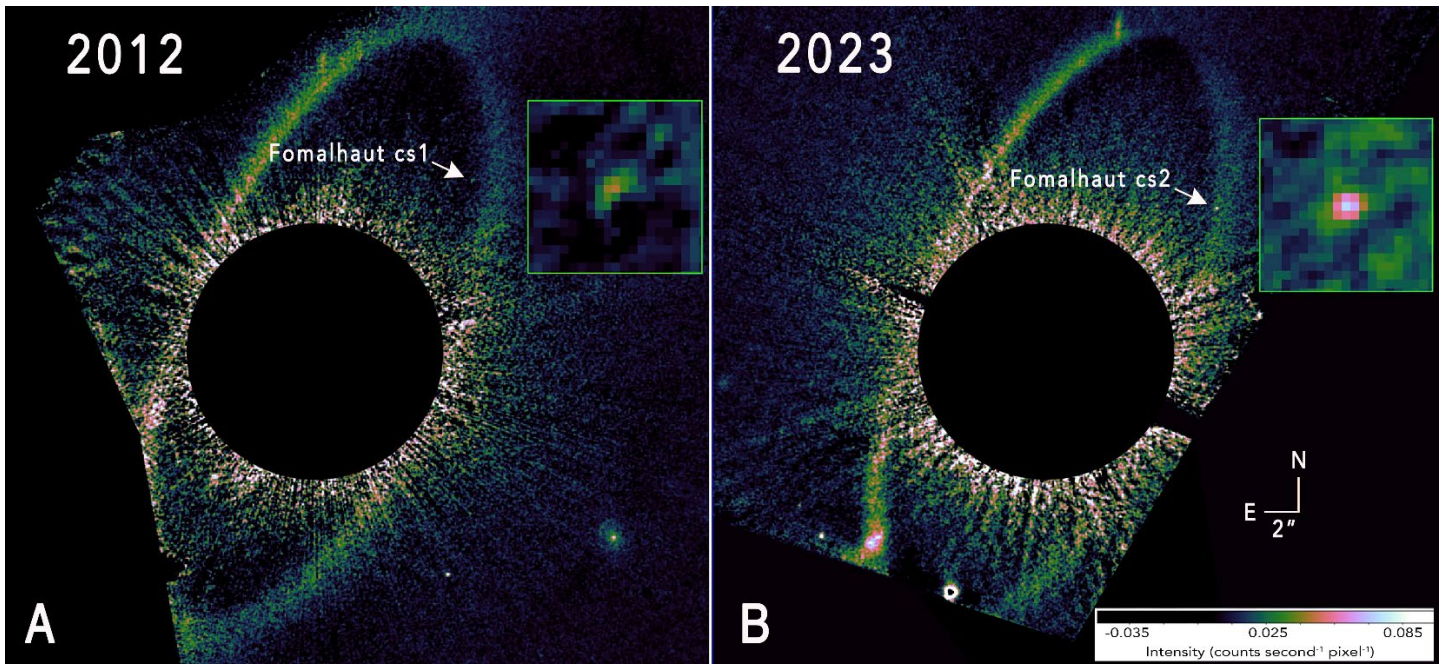

Fig. 1. Optical images of the Fomalhaut system in 2012 and 2023. Both panels are white light HST/STIS images. The bright central star has been artificially eclipsed with a coronagraph, to reveal the fainter belt-like dust structure around the star, with intensity indicated by the color bar. Black regions have no data, due to the coronagraph and the limited field of view of the camera. (**A**) Fom cs1 (white arrow) in our re-reduction of the 2012 observation (*12*). No other point sources are detected inside the dust belt. (**B**) Fom cs2 (white arrow) in the September 2023 observation, located northwest of the star at the inner edge of the dust belt. There is no conclusive evidence of Fom cs1 in 2023. In both panels, the inset boxes magnify a 1" × 1" (7.7 × 7.7 au) region centered on the identified point sources. All other point-like or extended sources in the field are known background stars or galaxies.

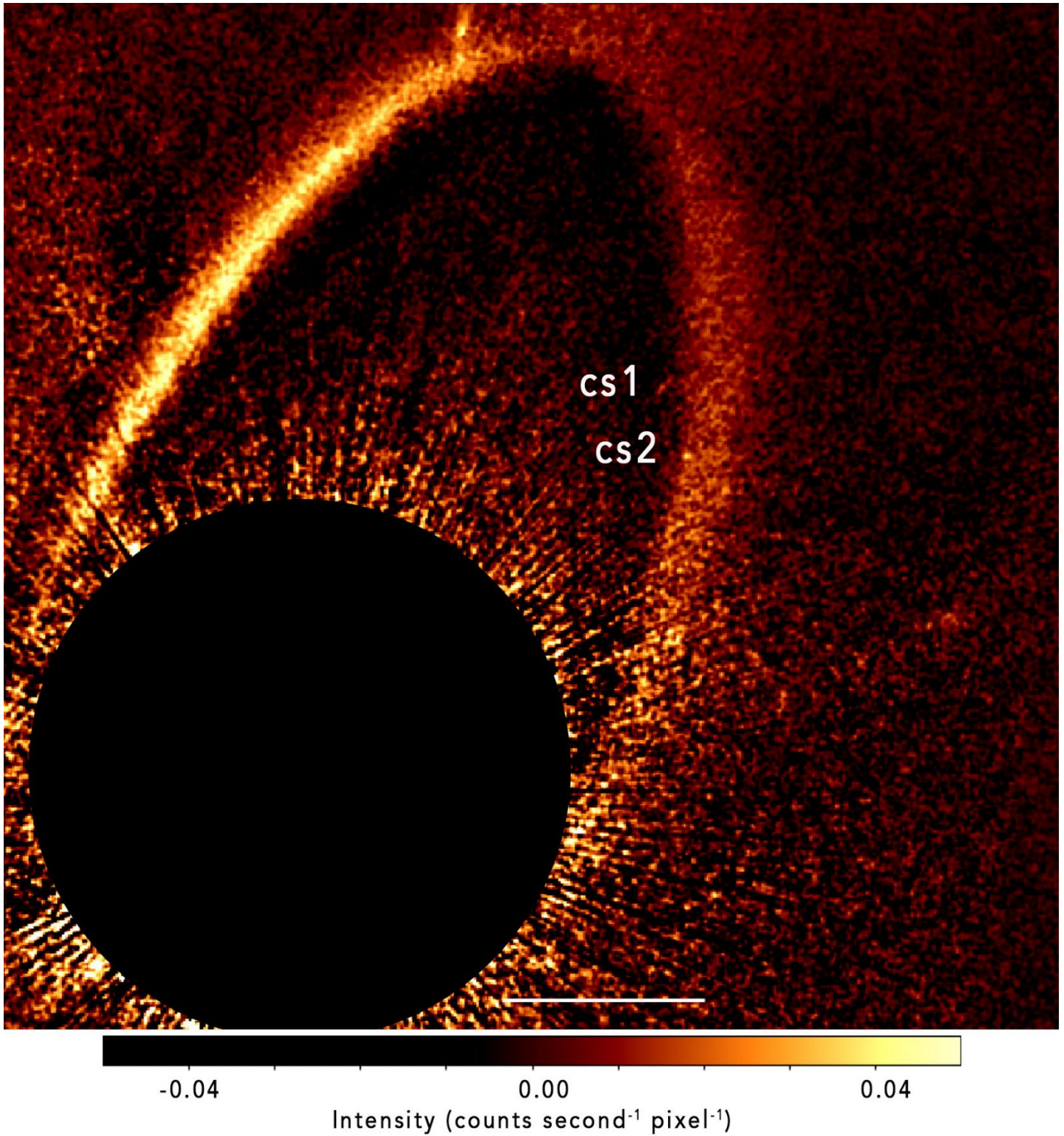

Fig. 2. Composite image of the 2012, 2013, and 2023 observations. The white light STIS images from three epochs have been averaged (intensity indicated by the color bar) to show the relative positions of cs1 and cs2 (white labels) a decade apart. Fom cs1 appears as a linear radial feature because its position moved between 2012 and 2013. The white scale bar is 5" (38.5 au). The data have been smoothed with a 0.8 pixel Gaussian filter. The dust belt is brighter to the east (left) because the dust grains have an asymmetric scattering phase function.

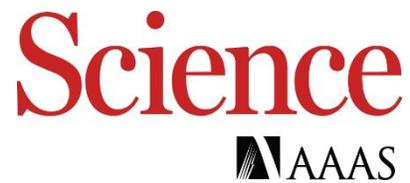

# Supplementary Materials for

**A second planetesimal collision in the Fomalhaut system**

Paul Kalas\*, Jason J. Wang, Maxwell A. Millar-Blanchaer, Bin B. Ren\*, Mark C. Wyatt, Grant M. Kennedy, Maximilian Sommer, Thomas M. Esposito, Robert J. De Rosa, Michael Fitzgerald

\*Corresponding authors. Email: kalas@berkeley.edu (P.K.), rbb@xmu.edu.cn (B.B.R.)

**The PDF file includes:**

Materials and Methods
Figs. S1 to S13
Tables S1 to S6
References *(37-53)*

# Materials and methods

1. <u>**2023–2024 observations and data reduction**</u>

**Table S1** summarizes the 2023 and 2024 Fomalhaut observations which were allocated 24 orbits. The HST/STIS instrument consists of a 1024 × 1024 pixel CCD with 0.05077" pixels yielding a field of view of 52" x 52". There are no filters available which means that the bandpass spans 0.20–1.03 μm (*11*). Bright stars can be artificially occulted in the focal plane. For most observations we used the horizontal occulting wedge at a location that has width 2.5" (WEDGEB2.5) and the gain is set to 4.015 electrons per data number (DN). The telescope's secondary support spider results in four bright diffraction spikes. The roll orientation of the telescope (ORIENTAT) is chosen so that the major axis of Fomalhaut's dust belt is not blocked by either the diffraction spikes or the occulting wedge. We did not use a subarray or dithering for this program. The point-spread function (PSF) subtraction strategy relies on angular differential imaging (ADI) (*37*) since the main objective is to detect point sources rather than characterize the dust belt. Each Visit in the program corresponds to a single orbit, and every orbit has a unique roll angle. To optimize sensitivity in the outer regions beyond 10" radius where Fomalhaut cs1 and the outer dust belt are located, exposure times were chosen in the range 25 to 30 seconds which lead to the saturation of the detector within 2–3" radius of the star. Cosmic rays and uncorrected hot pixels were identified and removed using a median combination of images given that the September 2023 data comprise a total of 429 individual exposures obtained across 18 unique telescope roll angles.

The first six orbits (Visits 01 to 06) failed to acquire a guide star and have no useful data (**Table S1**). The next 18 orbits (Visits 07 to 24) in September 2023 were mostly successful, with each orbit acquired at a different telescope roll angle to achieve PSF subtraction via ADI. The observations were grouped into three six-orbit blocks that were executed in consecutive orbits to minimize PSF variations over time from the changing thermal properties of the telescope. Visits 25 to 30 repeat the failed Visits 01 to 06, but only Visit 26 was successful. Visits 31 to 36 are a second attempt to repeat Visits 01 to 06, but only Visit 33 was successful. We found that the PSF of Visit 33 could be subtracted using the data from Visits 07 to 24. However, this was not the case for Visit 26 which had a PSF mismatched to all the other data because the occulting element used was WEDGE2.0 (width 2.0") instead of WEDGE2.5 (due to telescope roll and field of view constraints).

The multiple exposures in the flatfielded (_flt.fits) data files that were downloaded from the Mikulski Archive for Space Telescopes are individually processed. The first step is to correct pixels flagged with values 16 or 8192 in the data quality (DQ) file for each exposure. Each flagged pixel is replaced with the median value of neighboring pixels. The second step is to divide the entire image by the exposure time. Next the background sky value is determined (and subtracted) by taking the median value of pixels in the upper left corner of the STIS CCD which is the farthest point from the WEDGE2.5 occulting position. The fourth step is to implement a geometric distortion correction using the stistools package x2d (version 1.4.7) provided by STScI. The fifth step is to collapse the individual exposures into one file representing the cumulative integration of each orbit of observation. Various methods are tested at this stage, such as the median, the average, or the average with the rejection of high and low pixels in the image stack. The process with the best rejection of hot pixels and lowest noise is the average with the rejection of the highest pixel in the image stack. The sixth step is to co-align (register) all the orbits to a single fiducial orbit by iteratively shifting one image relative to the other to minimize the residual noise.

At this point the 18 images that each represent a single orbit are ready for three data reduction methods (DRM) conducted independently by three of the co-authors (see Author Contributions) to subtract the PSF. A fourth DRM by a fourth co-author independently followed the steps described above to produce a separate set of 18 images that were used for a fourth version of PSF subtraction (see below). This approach, which yields four separate versions of the PSF-subtracted data, allows estimates of the systematic uncertainties due to methodology differences, as well as an assessment of any spurious features that may contaminate the data. To determine the position of the star behind the occulting wedge of a single fiducial orbit, two techniques were used: drawing straight lines through the brightest pixels in the diffraction spikes and measuring their intersection (DRM1 and 4), or using the Radon transform (DRM2 and 3).

**Figure S1** shows the SNR maps for each version of the data reduction. The SNR maps are generated by taking the images after each PSF subtraction process described below, high-pass filtering to remove the extended circumstellar disk signal in the data, convolving the image with the instrumental PSF to smooth out pixel-to-pixel noise, and dividing each pixel by the standard deviation of counts in a 10-pixel-wide annulus centered about the star (*38*). The SNRs for Fom cs2 were determined by taking the maximum value of the SNR map near the location of Fom cs2.

**DRM1** produced two versions of the reference PSF. The first is a master PSF for each of the three six-orbit blocks shown in Table S1. The six orbits were averaged to make a master PSF but the two maximum pixel values and the single minimum pixel value was excluded from the average. The second is a pairwise PSF where a given orbit is PSF-subtracted 15 times by the other 15 orbits. The results reveal which pairs of orbits had the closest match in PSF structure and lowest noise in the difference image. For a given orbit, a pairwise PSF is constructed by taking the median of the results that achieved the lowest noise. Next, for a given orbit, the noise is measured in the two results that use the master PSF and the pairwise PSF, and the one that has the lowest noise is selected for further processing. The pairwise method was best for Visits 07 and 13, neither method was successful for Visits 19 and 20, which were excluded from further processing (representing 48 exposures of 30s each), and the remaining orbits attained the best PSF subtraction using the master PSF method. Each PSF-subtracted orbit is then rotated to a north-up orientation so that the astrophysical features are registered in every frame. The final step is to combine the PSF-subtracted, north-oriented orbits into a single image that represents the cumulative integration time for the September 2023 data (381 exposures of 30s each). Options for this step include taking median, average, or various rejection choices, such as to reject the high and low pixels, or to reject the orbits that display the highest residual noise. The final image with the lowest noise was the average that excluded the single highest and single lowest pixels (**Fig. S1A**). We were unable to obtain a satisfactory PSF subtraction for Visit 26 by using Visits 07–24 since the data were obtained at different wedge positions and hence different detector locations (**Table S1**). However, we achieved a satisfactory PSF subtraction using the pairwise method for Visit 33 (2024-09-10).

**DRM2** used the Karhunen-Loève Image Projection (KLIP; *39*) algorithm as implemented in the python package pyKLIP (*40*). The goal of these reductions was to detect and characterize any point sources in the images. In addition to the 2023 and 2024 epochs, the 4 previous epochs between 2010 and 2014 were processed as well. In all epochs, a Fourier-based high-pass filter was used to remove large-scale structures such as the debris disk. For all epochs except the 2024 epoch, ADI was used to disentangle the stellar PSF from any companions in the field. All images where the telescope rolled by at least 1 degree were used as reference images for KLIP, and 5 Karhunen-Loève (KL) modes were used to reconstruct and subtract off the stellar PSF. In the 2024 epoch, given there was only one roll angle, we used the 2023 images as reference stellar PSF images, essentially doing reference star PSF subtraction. 30 KL modes were constructed from the 2023 images and used to subtract the stellar PSF in the 2024 images. Fom cs1 was detected in the data between 2010 and 2013. While Fom cs1 was not seen in the 2014, 2023, and 2024 epochs, Fom cs2 is detected in the 2023 epoch (**Fig. S1B**), and tentatively detected in 2024. In all epochs, point sources in the reference images can distort the signal and thus the measured properties of any companions in the field of view. Analytical forward modeling of point sources through KLIP followed previous methods (*41*) to measure the astrometry of any point sources.

**DRM3:** To provide exact star centering between individual frames, a new PSF modeling algorithm was developed to minimize speckle residuals within a specified region (e.g. similar the Locally Optimized Combination of Images; *42*), while simultaneously optimizing for x (column) and y (row) image shift and a scaling coefficient for each reference image (allowed to be positive or negative). In each STIS dataset (obtained in 2010, 2012, 2013, 2014, 2023 and 2024), all the integrations for a single roll angle were combined into a single science image. In all years except 2024, a reference PSF library was constructed from images obtained at different roll angles. For the one roll angle in 2023, a library of PSFs from the 2023 Fomalhaut observations were used. For each science image a new set of image offsets and scaling coefficients were calculated for each image in the PSF library. For each science image a PSF-subtracted image was calculated by first shifting and scaling all the reference star images and taking their average. The offsets and scaling coefficients were obtained by minimizing the residuals in a region defined by a binary mask. The mask was created to avoid areas in the minimization where the detector is blocked by the various coronagraph masks, areas where the detector was saturated and areas with unusually high variance across the set of images, and areas outside of the field of interest (e.g. outside of the disk). This method detected Fom cs1 in 2012 and 2013, but in no subsequent datasets. Fom cs2 was detected in the 2023 epochs (**Fig. S1C**) and tentatively detected in 2024.

**DRM4** prepared the 2023 observation using the _flt.fits files from STIS for angular differential imaging reduction (ADI). First, for each science (SCI) extension of a _flt.fits file, we corrected charge transfer inefficiency and rectified the STIS CCD image following previous methods (*43*). Second, we used an X-marks-the-spot method to find the centers of the stars behind the STIS coronagraphic occulter: on each SCI extension, we fit Gaussian profiles to the peaks of the diffraction spikes, and then fit two straight lines to these peaks to find their intersection point, which is adopted as the location of the star. Third, we performed bad pixel correction for each SCI extension: the pixels that are flagged as bad pixels (i.e., 16, 256, 8192) in the corresponding DQ extension, were replaced by the median of their 5-by-5 neighboring non-bad pixels. With the prepared data, we shifted the star locations to the centers of the arrays and reduced the data in units of counts (ct) s$^{-1}$ pixel$^{-1}$. We first obtained an ADI template for the observations with a median of the prepared images. We then subtracted the median from the prepared observations, derotated the images to north-up and east-left and obtained the on-sky distribution of the residuals. We obtained a median of the derotated residuals while ignoring the pixels that were occulted by the coronagraphic occulters or the diffraction spikes. The final image (**Fig. S1D**) is north-up and east-left for analysis. We also used the 2023 template to reduce the 2024 dataset, and identified the candidate cs2 source in the 2024 observation. We repeated the reduction procedure for the 2023 data on the observations made in 2014 and earlier, but did not detect cs2 in these earlier observations.

As shown in **Fig. 1**, the limited field of view vignettes the southern portion of the belt. Due to the residual speckle halo surrounding the central star, portions of the belt are not detected to either side of the minor axis directions. We consider what fraction of the belt our data are sufficient to detect a cs2-like source near the inner edge of the belt. We implanted artificial PSF's scaled to the brightness of cs2 and determined that the northern wedge of detectability spans 68° from PA=10° clockwise to PA=302°. South of the star, a 13° detectability wedge extends between PA=168° to 181°. Southeast of the star the wedge of detectability is 15° between PA= 128° to 143°. Therefore, in the sky plane, a cs2-like object located at the inner edge of the dust belt would be detectable within a total of 96° of position angle. When the belt is deprojected assuming PA=336.2° and inclination 66° (*9*), the detectability wedges translate to 162° or 45% of full coverage. Fom cs1 and cs2 were discovered within 5.7° of each other in the sky plane, which translates to 8.1° in the deprojected belt plane.

## 2. 2023 cs2 astrometry, photometry and morphology

**Table S2** gives the measurements for the centroid locations of cs2 with respect to the star using four different final images independently processed by four authors. The astrometric uncertainty quoted in the last row is the standard deviation of the mean of these four measurements. This approach focuses on the systematic uncertainties arising from differences in data reduction methods (a similar methodology has previously been applied to $\beta$ Pic (*44*)) that typically exceed the 5 to 10 mas statistical uncertainties that we estimated from the data. We use additional systematic uncertainty estimates for HST/STIS [(*9*), their table 2] as 17, 5, and 1 mas for the geometric distortion correction, the stability of the optical distortion, and the detector position angle, respectively. This equals 18 mas uncertainty when added in quadrature. The last rows of **Tables S2** and **S3** show that the uncertainties arising from differences in data reduction technique are 11 to 19 mas for the right ascension (RA, $\alpha$) and declination (Dec, $\delta$) offsets. Adding these in quadrature with the systematic uncertainty gives uncertainties of (25, 21) mas for the 2023 (RA, Dec) offsets, and (26, 22) mas for the 2024 (RA, Dec) offsets relative to the star.

We conducted aperture photometry on cs2 for the 2023 data in a 5-pixel radius aperture on the four different versions of the data reduction. Following (*9*), we adopt a zero point of 24.48 mag (1 count $s^{-1}$= 24.48 mag) in the Vega magnitude system; 1 count $s^{-1}$ is equivalent to $4.628 \times 10^{-4}$ mJy. The local background is computed from the median of pixel values in an annulus between 5 and 10 pixels in radius. The mean and one standard deviation of the photometry is $24.67 \pm 0.15$ mag (24.75, 24.51, 24.58, 24.83 mag for DRM1, 2, 3, and 4, respectively). We performed the same photometry on cs1 in the 2012 data and found $24.98 \pm 0.11$ mag, in agreement with previous work (*10*). Therefore, cs2 is 0.3 mag brighter or a factor of 1.3.

Measurements of the full-width at half-maximum (FWHM) for cs2 in the various data sets (without high-pass filtering) yielded values in the range 2.1–3.0 pixels which are consistent with a point source superimposed on the faint nebulosity of the belt (see also **Section 6**).

## 3. Excluding cs2 as a stationary background object

If Fom cs2 were a stationary background object, it would appear at locations southeast of its 2023 position (relative to Fomalhaut) at earlier epochs given Fomalhaut's proper motion ($\mu$) of $\mu_\alpha$ = 328.95±0.50 mas yr$^{-1}$, $\mu_\delta$ = –164.67±0.35 mas yr$^{-1}$ (*45*). To rule out a stationary background object we carried out a process of injecting a source with similar brightness to Fom cs2 in previous epochs.

The brightness of Fom cs2 in the 2023 epoch was first determined by fitting the PSF using the TinyTim (version 7.5) software (*46*), to the high-pass filtered PSF-subtracted data (PSF-subtracted using DRM3 described above), using the Emcee MCMC sampling package (version 3.1.6) (*47*). Free parameters in the fit included the x and y positions of the PSF, the flux scaling parameter (a), and an additive noise term. Using this process, the flux scaling term was constrained to 10% ($\alpha$ = 3.3 ± 0.3), and the location constrained to ± 0.1 pixels.

Using these results, a flux-scaled TinyTim PSF was then injected into the 2012, 2013, and 2014 data (also reduced with DRM3) at the expected locations of a background object. The 2012 and 2013 data were then shifted such that all background objects lined up with the 2014 data. The co-aligned data from all three epochs were then combined using a weighted mean, where the weights were equal to the inverse of the variance of a 10×10 pixel box centered on the background object location in the data without an injected PSF.

**Figure S2** displays the co-added images with and without the injected PSF. In this image the PSF brightness was scaled to the 1-sigma lower limit of best-fit flux scaling range (i.e. $\alpha$ = 3.0). The injected PSF is detected above the residual noise, whereas no such point source is seen in the original data.

We note that the background field has very few stellar sources because Fomalhaut is well south of the galactic plane ($l$ = 20.5°, $b$ = –64.9°). Background stars that are as optically faint as cs2 are probably at great heliocentric distances in the galactic halo, supporting our assumption in the experiment above that a background source would be stationary. For example, if cs2 ($m_v$ = 24.7 mag) were an M dwarf (absolute visual magnitude $M_v$ = 8.8 mag), then the distance modulus is ~15 kpc. In **Fig. S2**, we co-added data separated by 2.3 years (May 2012 to September 2014). For a non-stationary object, the sharpness of the artificial PSF would smear and diminish in intensity if it moved by more than one pixel in this time period. (in **Section 6** we measure a FWHM = 3.6 ± 1.4 pix for a point source injected into the data.) For a source at 15 kpc, one pixel (0.05077") corresponds to 1.14 × 10$^{11}$ km. A star at 15 kpc would need a sky-plane motion of ~1500 km s$^{-1}$ to travel more than one pixel in 2.3 years. Even the fastest of the very rare high velocity stars known in the Galaxy move slower than this (*48*) supporting our assumption of a stationary background object.

### 4. A candidate cs2 source in the 2024 observation

The follow-up observations of Fomalhaut mostly failed due to technical problems with HST. Nevertheless, a single orbit acquired data on Universal Time (UT) 2024-09-10 and we found that its PSF could be subtracted using the data acquired a year earlier. **Figure S3** displays the four final PSF-subtracted images for these 2024 data and **Fig. S4** provides the SNR maps. In the reference frame of the star, a persistent feature appears ~2 pixels north of cs2's 2023 location. Though other pixels as bright as this feature appear elsewhere in each frame, none are fixed in location between the data reduction methods. For example, in **Fig. S3** the DRM1, 2, and 3 final images, which used a common set of calibrated flat-fielded images, show a single bright pixel offset two pixels to the upper left of the candidate 2024 cs2 source, but this same pixel does not have a high value in the final reduced image of DRM4, which independently processed the raw flat-fielded images.

We investigated if there was any step in the data reduction that was responsible for introducing the bright pixels which comprise the candidate source. We re-reduced the data multiple times where each version modified or excluded a procedure. For instance, a version of the processed data was created with the geometric distortion correction excluded. If the distortion correction requires the interpolation of pixel values, it could cause this specific spot on the detector to artificially enhance a group of pixel values. Or, in a different version, we do not correct the pixels flagged with the value 16 in the DQ file, and in another version we do not correct the pixels flagged with the value 8192, and in a third version the information from the DQ file is not used at all. Or, in other versions of the data reduction the 25 exposures that comprise Visit 33 are combined using the median, the average, or the average with various combinations of rejection parameters, such as rejecting the two highest and two lowest pixel values for a given pixel location. These steps are effective in rejecting cosmic rays. However, none of these experiments resulted in a final reduced image where the 2024 candidate cs2 source was absent.

**Table S3** gives the astrometric measurements for the cs2 candidate in 2024. The northward motion is in the same direction as cs1 in the 2004–2006 HST observations. Assuming the 2024 source is astrophysical, **Fig. S5** and **Table S4** give the orbital elements fitted using the Orbitize! package (*49*) and its implementation of the orbits for the impatient (OFTI) algorithm (*50*). The mean and one standard deviation of 3-pix radius aperture photometry on the candidate source gives $24.68 \pm 0.23$ mag (24.83, 24.46, 24.50, 24.93 mag for DRM1, 2, 3, and 4, respectively) which is consistent with the 2023 photometric measurements of cs2 (see **Section S2**).

5. **Search for cs1 in 2023**

The observations were designed to search for cs1 by doubling the number of orbits (24) compared to prior epochs (12). In the region >10" radius from the star the noise decreases as the square root of the integration time, so we expected to improve the sensitivity by a factor of 1.4. Unfortunately, the failure of six orbits provided only 18 orbits of data which leads to a factor of 1.2 improvement over prior epochs. All else being equal, Fom cs1 would likely be detectable even if its brightness decreased by 20%.

However, one major difference between 2023 and the previous deep imaging epoch of 2013 is that cs1 may overlap with the nebulosity of the outer dust belt in the sky plane. Therefore, the photon noise from the nebulosity adds an additional source of noise in 2023 that was not a factor in prior epochs. An additional uncertainty is predicting where Fomalhaut cs1 should be expected in the belt, or if it crossed to the outer fainter edge of the belt in 10 years. If cs1 stayed on the radial trajectory indicated by the 2012 and 2013 epochs of astrometry and maintains either a constant velocity, or accelerates due to radiation pressure, then in 2023 cs1 would either be superimposed on the dust belt or travel beyond the outer boundary in the acceleration case.

Using a dust cloud model, previous work (*13*) predicted that the stellocentric location of Fom cs1 on UT 2022-10-22 would be $(\Delta\alpha, \Delta\delta) = (-9377, 11144)$ mas for a bound orbit and $(\Delta\alpha, \Delta\delta) = (-9809, 11665)$ mas for an unbound orbit. These values correspond to 14.564" separation at PA = 319.9° for the bound orbit and in 15.241" separation at PA = 319.9° for the unbound orbit.

However, in Section 6 we estimate cs1's acceleration between 2012 and 2013 as $1.9455\times10^{-7}$ km s$^{-2}$ in the radial direction. This would place it farther away from the star in 2023 compared to the previous prediction (*12*). To estimate the distance traveled between 2013 and 2023, we adopt a previous (*51*) asymptotic limit to the velocity of eccentric grains accelerated by radiation pressure: $v_\infty = v_i \times \sqrt{(2\beta - 1 + e)}$, where $v_i$ is the initial velocity, $e$ is the initial eccentricity of the parent bodies, and $\beta$ is the overpressure force (the ratio of stellar radiation pressure to the gravitational force) which is constant with radius from the star. The $\beta$ value for 0.1 $\mu$m grains orbiting A stars like Fomalhaut is 20 to 30 (*27*), depending on the grain properties. If we assume the initial velocity of grains is the velocity of cs1 from the 2004 and 2006 detections (~4 km s$^{-1}$; **Table S6**) and $e \sim 0.1$ (*5*), then for $\beta = 30$, $v_\infty = 4 \times \sqrt{(2\times30 - 1 + 0.1)} = 31$ km s$^{-1}$ and for $\beta = 20$, $v_\infty = 25$ km s$^{-1}$. We adopt the latter value.

The velocity component (which is entirely in the radial direction) in the deprojected plane between the 2012 and 2013 epochs is 11.682 km s$^{-1}$ (**Table S6**). We consider when and where the cs1 grains reach the asymptotic value of 25 km s$^{-1}$, which we adopt as $v_\infty$. It takes $6.836\times10^7$ seconds (2.17 yr) for the velocity to increase by an additional $\Delta v = 13.3$ km s$^{-1}$ given the acceleration of $1.9455\times10^{-7}$ km s$^{-2}$. In this time, it traveled distance $d = v_i \times t + \frac{1}{2} \times a \times t^2 = 11.682 \times 6.836\times10^7 + 0.5 \times 1.9455\times10^{-7} \times (6.836\times10^7)^2 = 1.55\times10^9$ km = 8.4 au. July 2015 is 2.17 years after the last HST detection of May 2013 and Fom cs1 would be 8.4 au away from its 2013 position. For the additional ~8 years from 2015 until the 2023 HST observations, the grains traveled at their terminal velocity of 25 km s$^{-1}$, which corresponds to 8 yr × $3\times10^7$ s yr$^{-1}$ × 25 km s$^{-1}$ = $6\times10^9$ km $\cong$ 40 au. In summary, in 2023 we expect the 0.1 $\mu$m grains to reside in a radial direction away from the star that is an additional ~48 au farther from the 2013 location at 128 au in the belt plane. In the sky plane this corresponds to a stellocentric separation of 18.8" at PA = 318.3° ($\Delta\alpha = -12.512"$, $\Delta\delta = 14.044"$).

In **Fig. S6** we show a magnified view of our 2023 data presented in **Fig. 1** after high-pass filtering the data to remove the diffuse nebulosity from the belt. Based on the previous discussion, we define a search wedge by two dashed lines that are 20" long and oriented 2° to either side of cs1's 318° position angle. Fom cs2 is detected in these data but no compact sources are evident within our search wedge for cs1.

To better assess our sensitivity to Fom cs1, we implanted 14 artificial point sources radially along the search wedge. We reduced their flux by the amount appropriate for their inverse distance squared from the star relative to cs1. For example, if cs1 has moved 50 au away from its radial location in 2013, it has roughly half the brightness in 2023. The first implant is located at 14.5" separation and PA=318.4° in the sky plane (135 au in the belt plane), which is on the inner edge of the debris belt similar to cs2's location. This is just below the region predicted by (*13*) for the location of cs1 in a bound orbit. The third implant is located at 15.3" separation and PA=318.4° in the sky plane (143 au in the belt plane) which is just below the region predicted by (*13*) for the location of cs1 in an unbound orbit.

The results of the implant experiment (**Fig. S7**) show that Fom cs1 is difficult to detect in the absence of either accurate a priori information that can narrow the search region, or observations at a second epoch within 1–2 years to detect its motion. At least five of the implants are indistinguishable from speckle noise. Therefore, the recovery of Fom cs1 in these data is inconclusive, for three main reasons: i) 10 years elapsed since HST last observed Fomalhaut with deep observations, making it difficult to accurately predict cs1's current radial location from the star, ii) after 10 years the radial trajectory with acceleration may have increased cs1's stellocentric distance by 50 au with the corresponding ~50% reduction in reflected stellar light, and iii) 25% of our observations failed due to guide star acquisition issues so we did not achieve our planned sensitivity.

**Figure S8** shows a candidate source within our search wedge that appears to persist in all versions of our data reduction. It is located at separation 19.40" and PA=318.0° ($\Delta\alpha$, $\Delta\delta$= −12.97", 14.44") and has the low SNR~2–3 expected from the implant experiment. Aperture photometry carried out in exactly the same way as for cs1 in 2013 (see the next section) shows that this feature is ~1.0 mag fainter than cs1 in 2013. This is 40% of cs1's flux in 2013 which is consistent with the 49% value that we expect from its greater distance from the star. We cannot confirm whether this feature is an astrophysical source.

## 6. <u>Reanalysis of Fom cs1 in the 2013 data</u>

A prior analysis of the multi-epoch HST data (*10*) concluded that cs1 in 2013 had accelerated significantly towards the radial direction, its brightness decreased, and its morphology became extended in one year since the 2012 epoch.

This is unlike the previous four epochs of astrometry (2004, 2006, 2010, and 2012), which are consistent with a uniform motion model at 4.36 km s$^{-1}$ in a high-eccentricity ($e \sim 0.8$) orbit that is not radial (*9*). Fom cs1 appeared extended in the radial direction for the 2012 data, which is expected as a data reduction artifact. After PSF subtraction, residual correlated noise has the form of positive and negative radial diffraction spikes 2–3 pixels in width. The PSF wings of faint sources such as Fom cs1 can have their morphology altered to appear extended in the radial direction if the sources fall on a positive residual of this noise. Unfortunately, the radiation pressure blowout of small grains is also directed in the radial direction and higher signal-to-noise or multiple independent observations are required to confirm whether cs1 is elongated due to instrumental or astrophysical reasons.

We re-reduced the 2013 observations using the DRM1 and 2 methods (**Fig. S9**). Fomalhaut cs1 is detected with SNR = 4.0 and 3.6 respectively (**Fig. S10**). We analyzed the photometry, astrometry, and morphology in these two versions of the data reduction to compare against the previous results (*10*).

Fomalhaut cs1 is observed as a point source for both versions of the PSF subtraction. First, we measure the FWHM of a background star (14.4" separation, PA=212.9°) by fitting a Gaussian to the radial profile. We find FWHM = 2.2 pix and using the same fitting method on cs1 gives FWHM = 2.7 pix. This indicates Fom cs1 may be slightly extended relative to a background star. However, this background star is ~10× brighter than Fom cs1 and its radial profile is less susceptible to noise. To test the effects of noise on the FWHM measurements of Fom cs1, we created an artificial PSF using the TinyTim software package (*46*) and scaled its brightness to match the detection of Fom cs1. We then implanted the artificial PSF at 25 locations near Fom cs1. Six of these implants were not detectable because they were in regions of negative noise. One could not be fitted by a Gaussian function because it landed near a bright pixel, producing a double-lobed structure. The remaining 18 implants had a mean FWHM = 3.6 ± 1.4 pix (median = 3.9 pix). Therefore, the measured 2.7 pix FWHM of Fom cs1 is consistent with the expected properties of a faint point source in the 2013 observations. Our implant experiment shows that its measured FWHM varies substantially due to the residual noise and its mean value is generally greater than a brighter comparison background star due to the low SNR of the Fom cs1 detection.

The next steps are to compare our photometry and astrometry to the previous results (*10*). We conducted aperture photometry with a zero point of 24.48 mag as discussed above and using an aperture with radius 5 pix. The local background is computed from the median of pixel values in an annulus from 5 to 10 pixels radius. From five different versions of the PSF subtraction for the 2013 epoch, the mean and one standard deviation of the photometry is 25.00 ± 0.07 mag. An additional method for estimating the photometric uncertainty due to the noise is to perform photometry on the artificial PSFs that were implanted in the reduced data. These measurements give an 0.30 mag uncertainty. Combining the 0.07 and 0.30 mag standard deviations in quadrature gives 0.31 mag uncertainty in the photometry.

We now compare the 2013 photometry with prior epochs and with the previous measurements (*10*) which found Fom cs1's flux was 0.47±0.08 ct s$^{-1}$ for 2013. Our photometry of 25.00 ± 0.31 mag corresponds to $0.62^{+0.20}_{-0.15}$ cts s$^{-1}$, which is only $1\sigma$ brighter than the previous result (*10*), so the two are consistent. We conducted exactly the same photometry on our 2012 images of cs1 and found 24.98±0.11 mag. Therefore, the flux appears constant between 2012 and 2013. The previous 2010 photometry was 0.77±0.10 cts s$^{-1}$ (*10*) and 0.49±0.14 cts s$^{-1}$ (*9*). In summary, our 2013 photometry using the same instrument and similar data reduction methods do not detect any changes in Fom cs1's brightness in the September 2010 to May 2013 time interval.

Finally, we perform astrometry on our images for the 2013 epoch. **Figures S11–S13** display our results with respect to prior work. We confirm the previous finding (*10*) that Fom cs1 accelerated in a single year between the 2012 and 2013 epochs. This is visible by eye in **Fig. S11** since the one-year position displacement between 2012 and 2013 appears roughly equal to the two-year displacements in the 2004 to 2006 and 2010 to 2012 time periods. Assuming that Fom cs1's motion lies within the belt plane, the deprojection from the sky plane to the face-on view reveals a substantial change in the magnitude and direction of Fom cs1's velocity between 2012 and 2013 compared to the previous epochs.

Fom cs1's motion is consistent with a model of a dust cloud that is accelerated by radiation pressure (*10*). **Figure S13** plots the belt-plane positions of cs1 using the data from **Table S6**. The last column of **Table S6** gives the radial component of velocity in km s$^{-1}$. We therefore compute the radial acceleration between 2010-2012 and 2012-2013 as (11.682 - 3.354) (km s$^{-1}$) / 42806016 s = $1.9455 \times 10^{-7}$ km s$^{-2}$. This acceleration was used above to estimate the possible position of cs1 in the sky plane in our 2023 data set.

To summarize, our reanalysis supports the dust cloud hypothesis (*10*) based on Fom cs1's radial acceleration between 2012 and 2013, which is higher than is possible for a Keplerian orbit. However, contrary to previous work (*10*), we do not detect an extended morphology or dimming in flux for cs1.

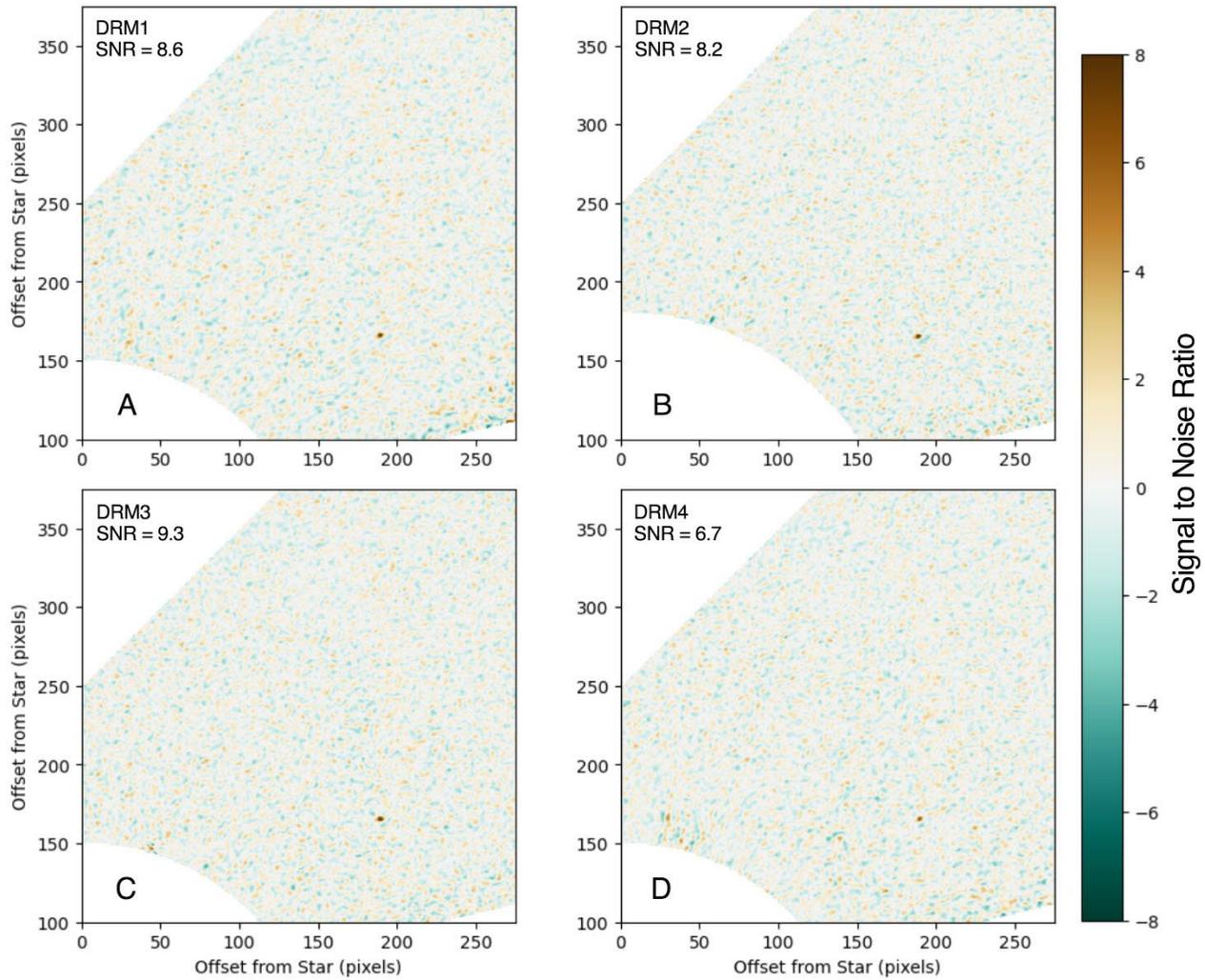

**Fig. S1**. **SNR maps of the four versions of data reduction.** Each panel is cropped to the region of interest containing Fom cs1 and cs2. The color bar ranges from SNR of -8 to 8, showing that there are no other features that are such large deviations from 0 in the SNR maps. Fom cs2 is detected in each data reduction method, with SNR between 6.7 and 9.3.

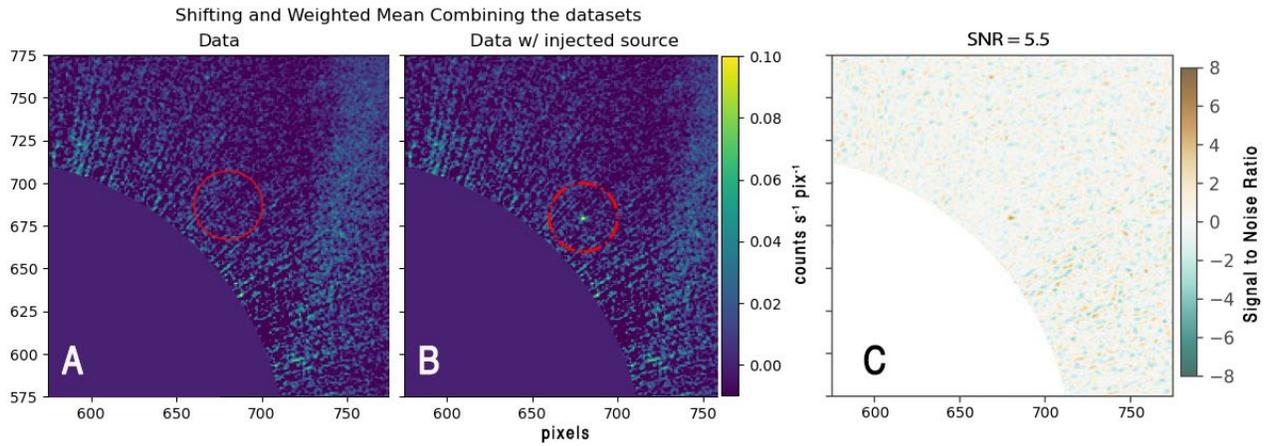

**Fig. S2**. **Fom cs2 is not a background star.** **(A)** Combination of the 2012, 2013 and 2014 HST observations co-aligned to background stationary objects. The center of the red circle (radius 1") shows its expected location at prior epochs if it were a background star. Due to Fomalhaut's proper motion to the SE, the location is closer to Fomalhaut than in 2023. **(B)** We inject a point source with the brightness of Fomalhaut cs2 to show that it would have been detected in these data. **(C)** Signal-to-noise map shows that the injected point source has SNR = 5.5.

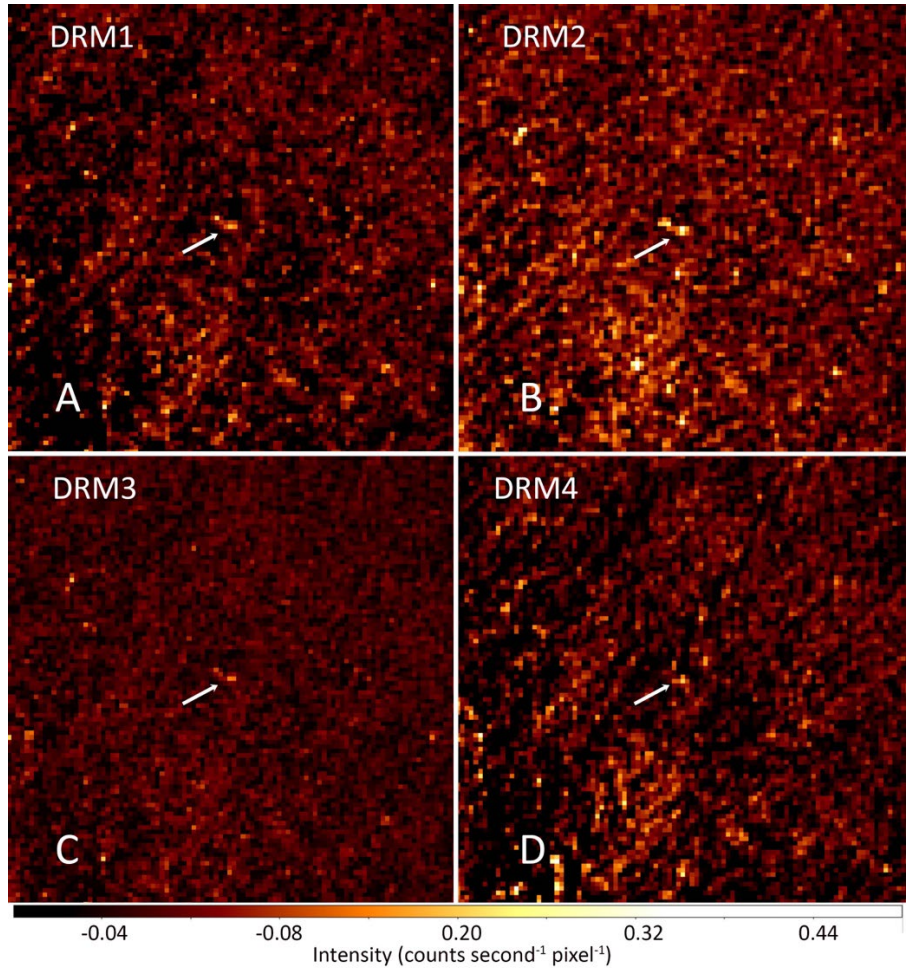

**Fig. S3. Four images of the single-orbit 2024 data.** PSF-subtraction was achieved independently with DRM1, 2, 3, and 4 in Panels A, B, C, and D, respectively. The white arrow points to a group of bright pixels that are located at the same stellocentric position in every version of the data reduction. Other bright pixels are evident in each frame but do not repeat their locations from image to image. The frames have dimension 100 × 100 pix (5.08" × 5.08") and north is up, east is left.

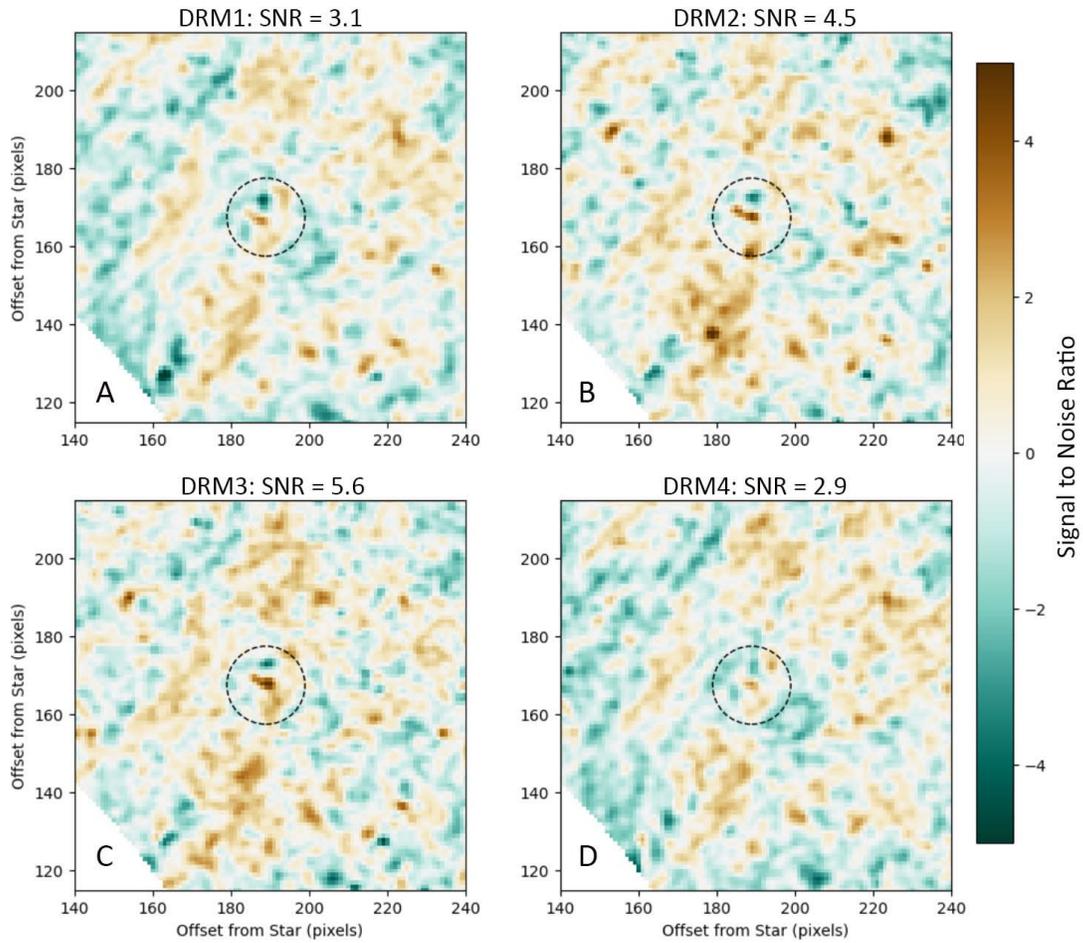

**Fig. S4. SNR maps for the 2024 data shown in Fig. S3.** The candidate source, which is located north of the 2023 position, is a feature that has SNR ranging between 2.9 and 5.6.

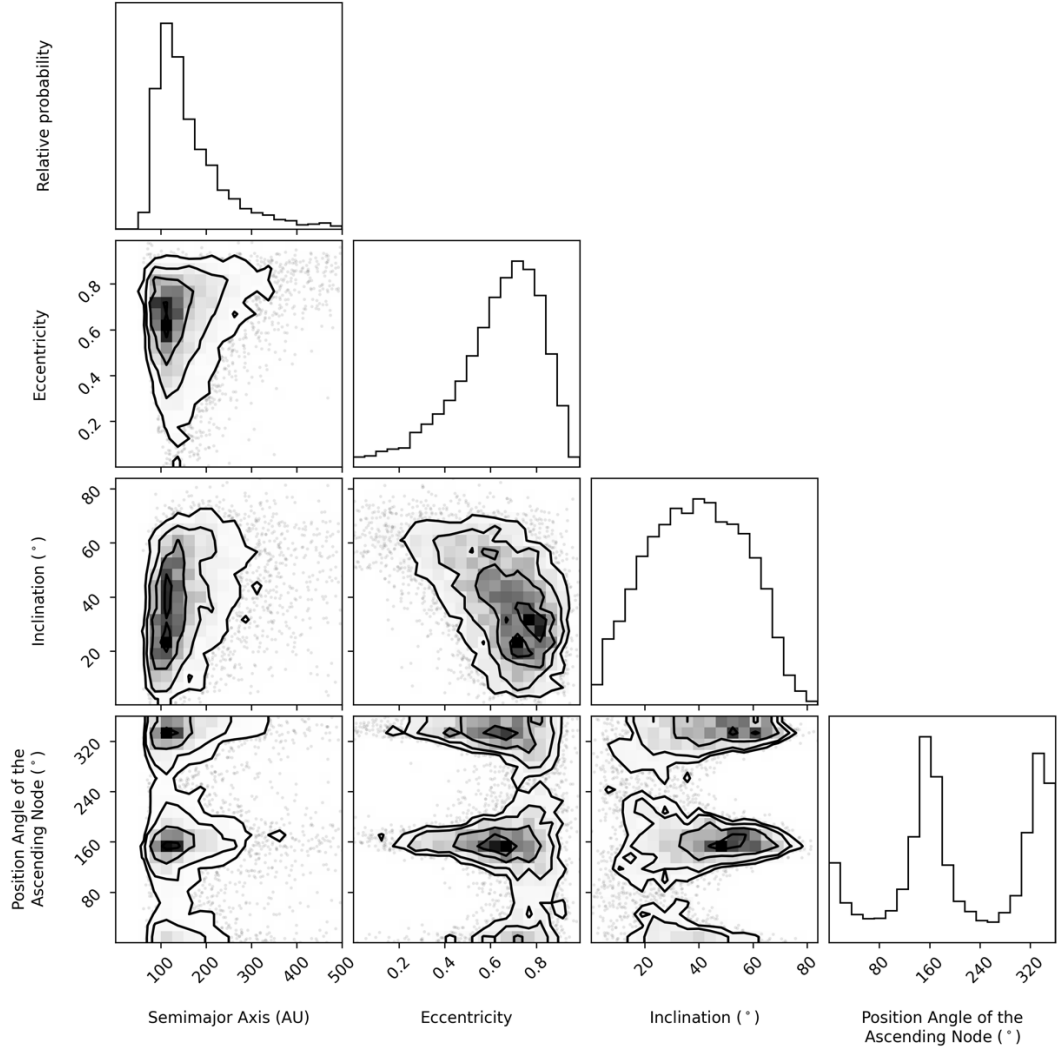

**Fig. S5. Posterior distribution of possible orbits for Fom cs2.** Along the diagonal, each panel shows a histogram of the 1-D posterior distribution of the orbital parameter corresponding to that row and column. The other panels contain 2-D posterior distributions that show the correlation between the two orbital elements corresponding to that row and column. The darker shaded regions represent areas of higher posterior probability density, and the contours correspond to the 11.8-, 39.3-, 67.5-, and 86.4-percentiles (corresponding to the 0.5-, 1-, 1.5-, and 2-sigma confidence intervals for a 2-D Gaussian distribution). The individual dots correspond to possible orbits that fall outside the 86.4-percentile distribution.

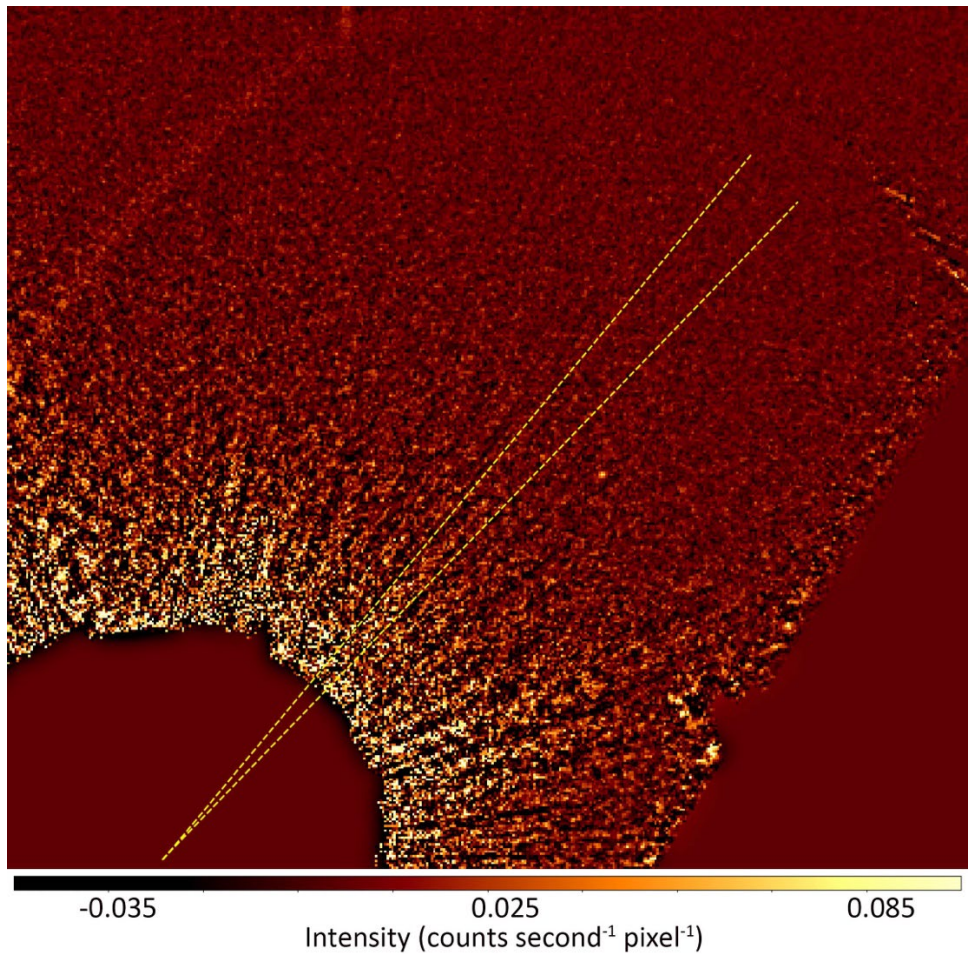

**Fig. S6**. **The 2023 high-pass filtered data.** Fom cs2 is detected but no other sources are evident in our Fom cs1 search wedge between the two dashed lines, which are 20" in length and oriented at PA=316° and PA=320°.

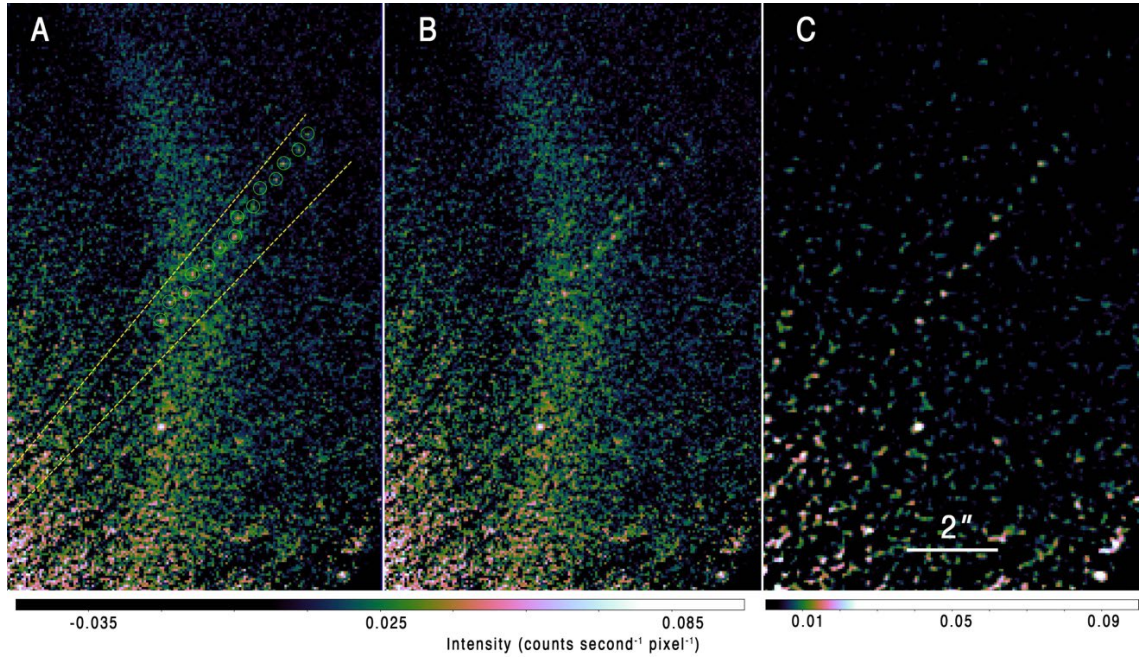

**Fig. S7. Fom cs1 implant experiment.** Fourteen artificial point sources are implanted along our search wedge in the 2023 data with their brightness reduced compared to Fom cs1 according to their greater distance from the star. Panels (**A**) and (**B**) are exactly the same except (**A**) marks the artificial sources with circles. Panel (**C**) high-pass filters the data to remove the nebulosity from the debris belt.

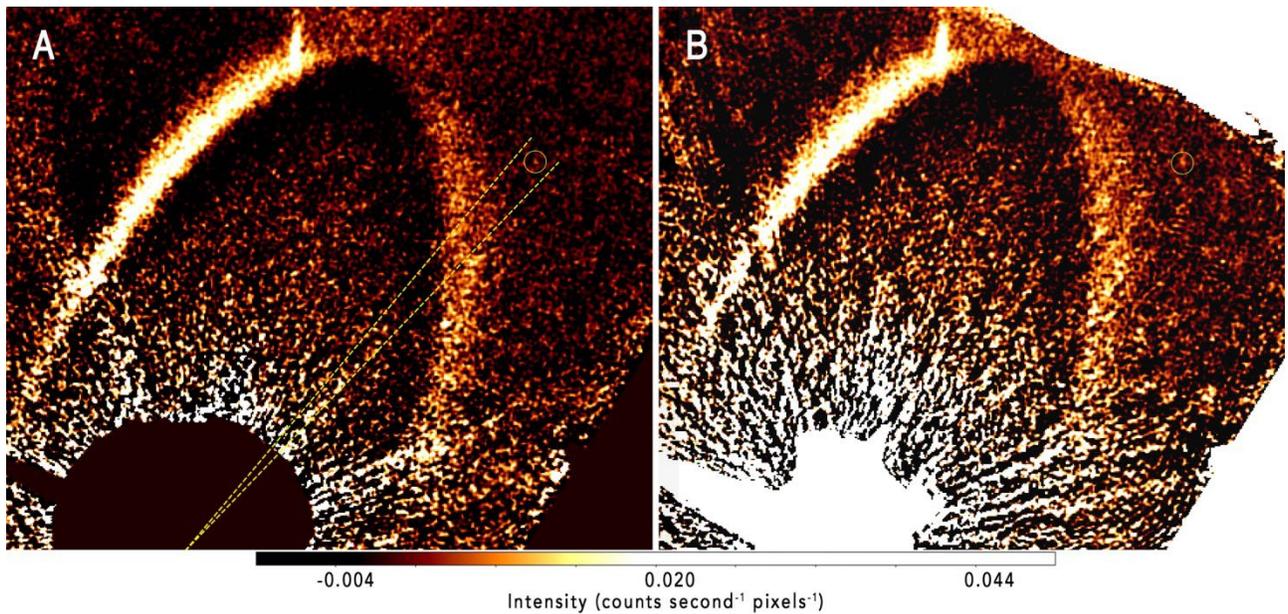

**Fig. S8**. **A candidate feature for the location of cs1 in 2023.** A 0.8" diameter circle marks the single feature within our search wedge that persists in all versions of the data reduction. (**A**) DRM3 and (**B**) DRM4 versions of the 2023 data reductions smoothed with a Gaussian filter (s = 1 pixel). The search wedge is the same as with **Fig. S6**. We cannot determine whether this source is astrophysical.

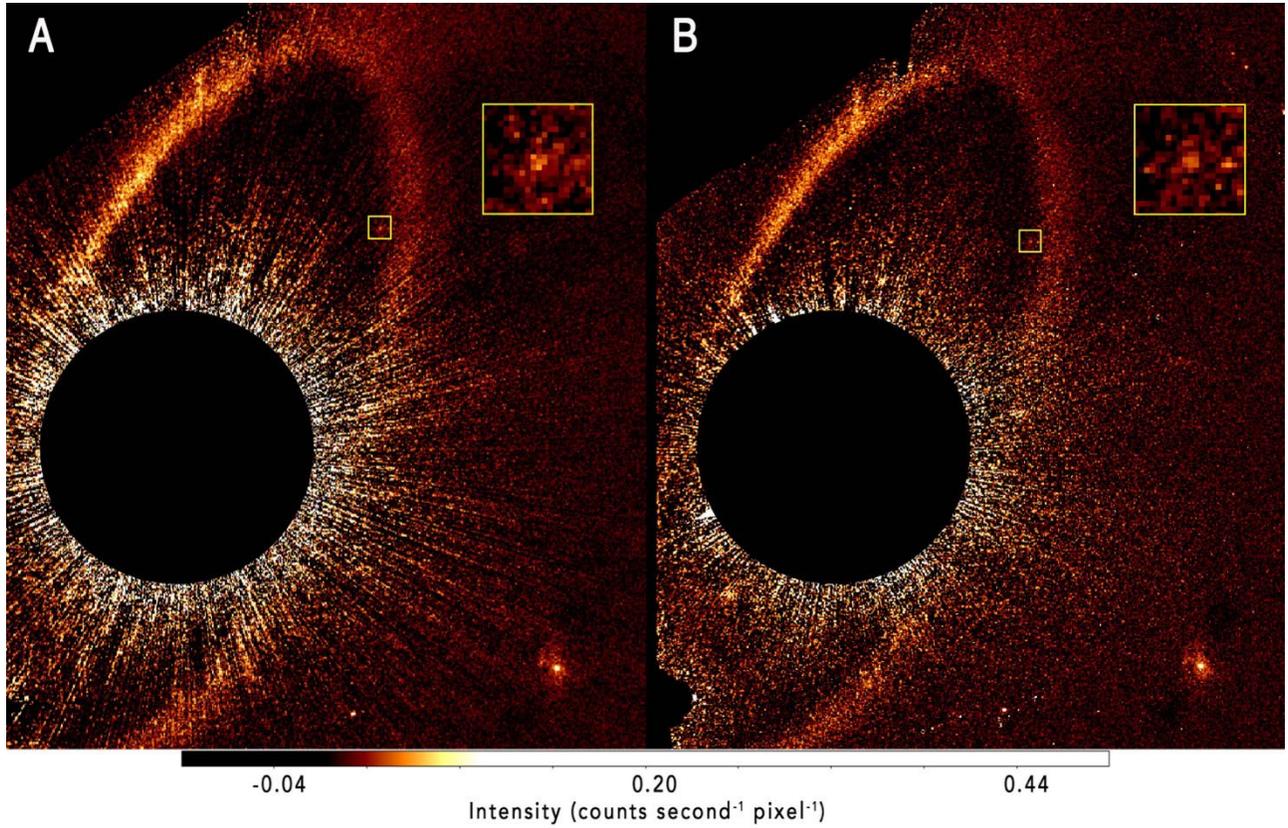

**Fig. S9. Two independently reduced versions of the 2013 data.** DRM1 and 2 are shown in Panels A and B, respectively. The Fomalhaut cs1 detections (yellow boxes) are consistent with a low SNR point source. The inset boxes magnify a 1"×1" region centered on cs1. All other point-like sources in the field are known background stars and galaxies.

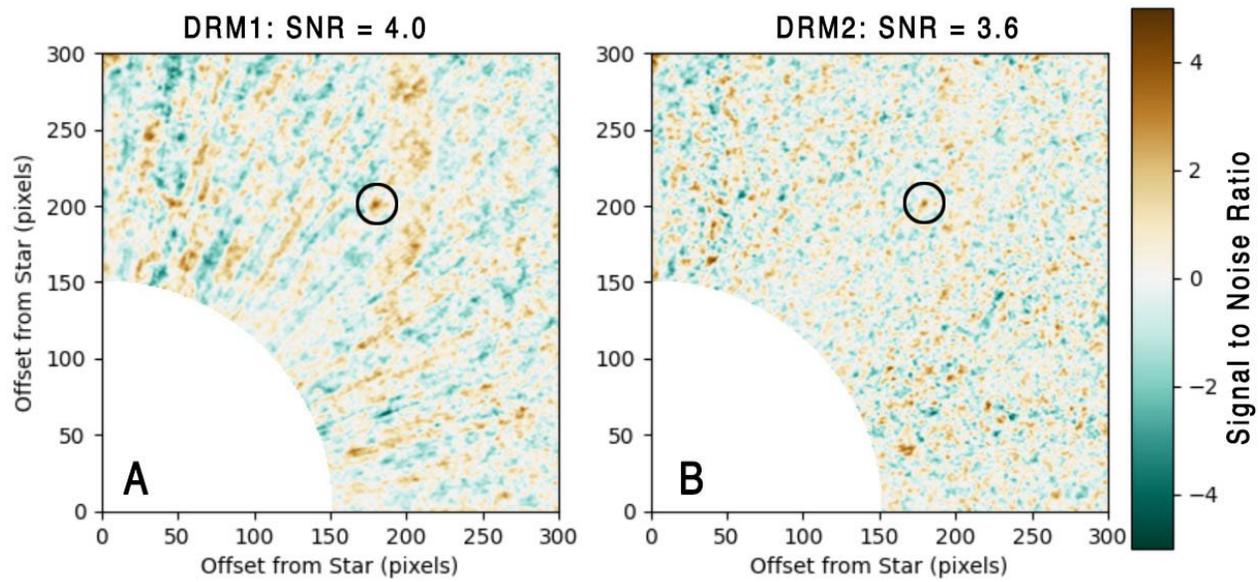

**Fig. S10**. **SNR maps for the 2013 data reductions in Fig. S9.** The detections of Fom cs1 have SNR values 4.0 and 3.6 in Panels A and B, respectively. The method for calculating the SNR is the same as for **Fig. S1**.

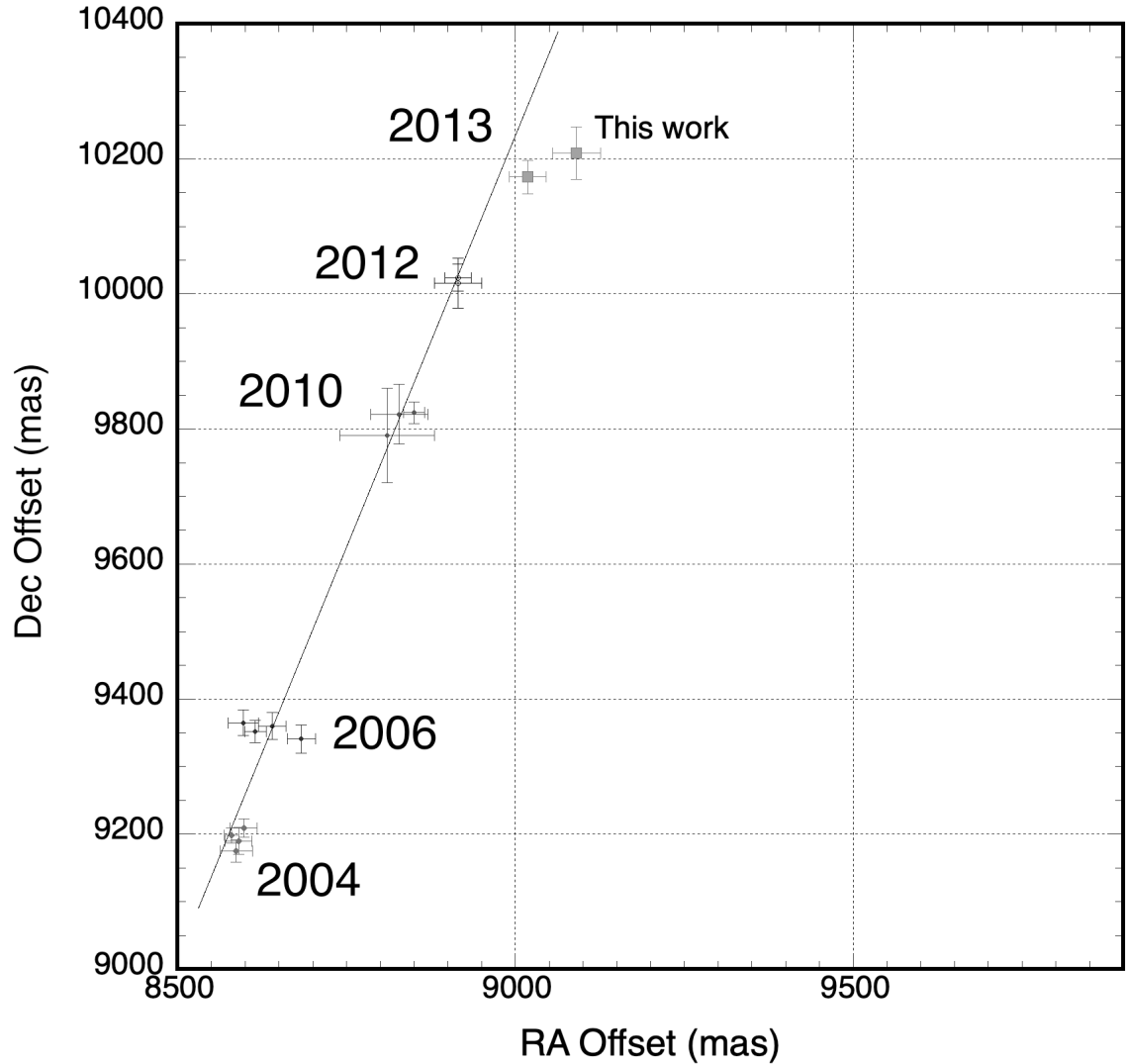

**Fig. S11. Astrometry of Fom cs1 relative to the star**. The 2004 and 2006 positions (*9, 10, 52, 53*) agree on the Dec offset. However, in 2006 the spread along RA is 86 mas, which is discrepant by 4-$\sigma$ (the published uncertainties are ~20 mas). We ascribe this disagreement to differences in analysis methods. The three points for 2010 (*9, 10, 53*) agree within the uncertainties. The 2012 points (*9,10*), are in agreement. The 2013 astrometry shows our re-reduction and previous results (*10*). Both results find a westward deviation compared to a straight-line fit though the 2004–2012 data. The 2013 points are also too far forward compared to uniform motion (*10*).

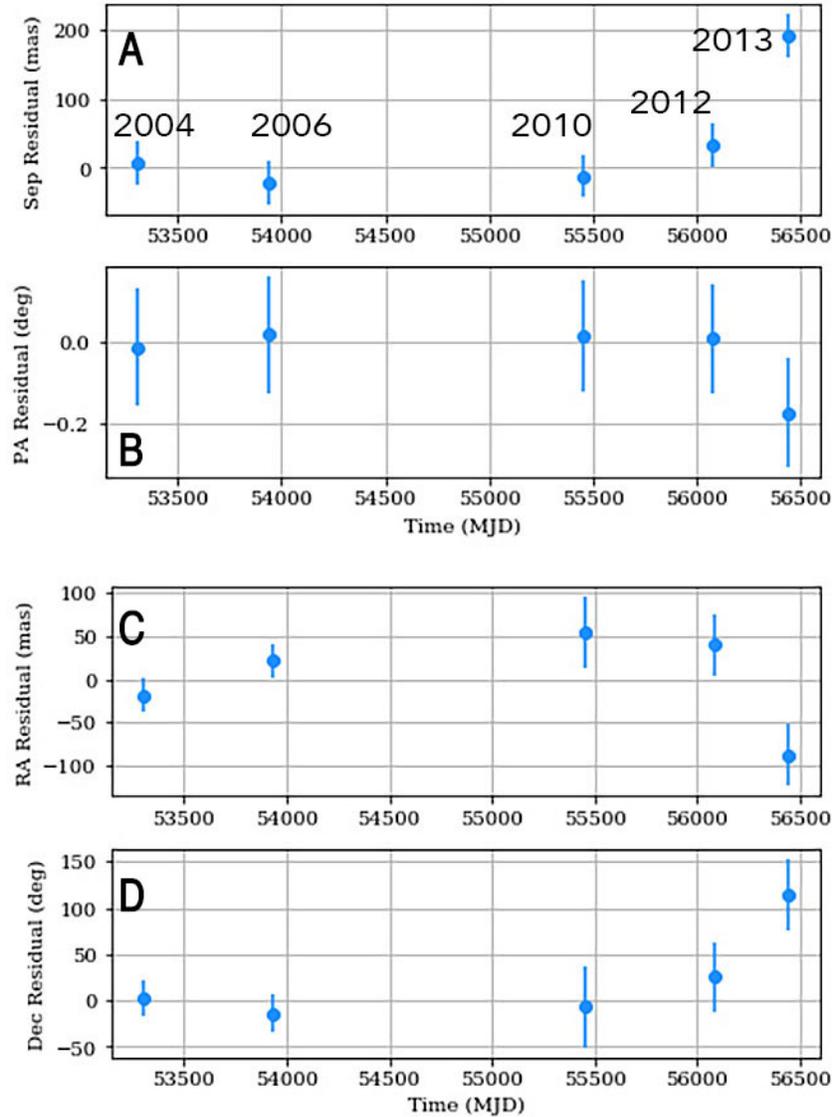

**Fig. S12. Astrometric residuals to the Keplerian orbit fit for Fom cs1.** The four panels show the astrometric measurement residuals as a function of time after subtracting off the best-fitting Keplerian orbit to the five epochs of astrometry between 2004 and 2013. Each panel shows the residuals along a different principal axis for determining the astrometry: radial separation (A), position angle relative to the star (B), right ascension offset from the star (C), and declination offset from the star (D). The error bars show the 1-sigma uncertainties for each measurement along that principal axis. Our measured position in 2013 is a 6-$\sigma$ outlier (~200 mas or ~4 STIS pixels) from our best-fit Keplerian orbit.

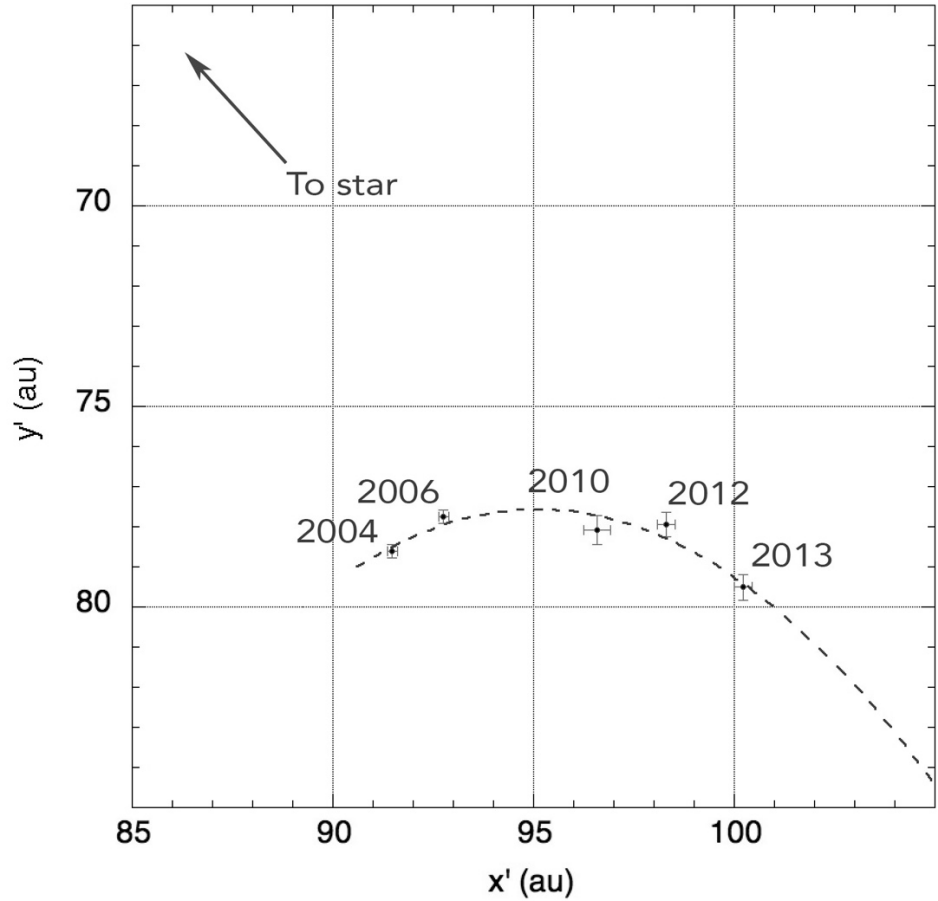

**Fig. S13. The deprojected belt plane positions from Table S6.** We assume cs1 orbits in the belt plane with $i = 66°$ and PA $= 336.2°$ (*9*). The star's location is at (0,0) and the belt's northwest semi-major axis is rotated clockwise to become the positive x axis [see Fig. 13 in (*9*)]. The conversion from arcseconds to au assumes $d$=7.704 pc and 0.05077"/pix. The dashed line is a least-squares 2nd order polynomial fit ($y = 714.61 - 13.394\,x + 0.070405\,x^2$).

**Table S1: Fomalhaut 2023–2024 HST/STIS observations.** All observations were made by occulting the star behind the WEDGEB2.5 position except for Visit 26 on 2023-11-14 that used the WEDGEB2.0 position. Each Visit corresponds to one HST orbit. The approximate midpoint of the 2023-09 observations is Julian Date JD 2460206.5 and for the 2024-09-10 observations it is JD 2460564.2.

| UT Dates | Visit | ORIENTAT (deg) | Exp Time (s) | # exp | Notes |
|---|---|---|---|---|---|
| 2013-09-14 | 01–06 | N/A | N/A | N/A | All failed |
| 2023-09-15 | 07–12 | -24.3, -31.7, -19.4, -29.3, -21.8, -26.8 | 30 | 144 | |
| 2023-09-19 | 13–18 | -23.7, -31.1, -18.8, -28.6, -21.3, -26.2 | 30 | 142 | 1st exp in Visits 14 & 18 failed |
| 2023-09-21 | 19–24 | -23.1, -30.5, -18.1, -28.0, -20.6, -25.6 | 30 | 143 | 1st exp in Visit 23 failed |
| 2023-11-14 | 25–30 | N/A, 14.1, N/A, N/A, N/A, N/A | 24 | 25 | Only Visit 26 did not fail |
| 2024-09-10 | 31–36 | N/A, N/A, -29.9, N/A, N/A | 25 | 25 | Only Visit 33 did not fail |

**Table S2: 2023 Fomalhaut cs2 astrometry.** RA and Dec values are offsets relative to the star assuming 0.05077"/pixel. The offsets in the plane of the belt are x' and y'. The original belt observed in the sky plane in the x,y Cartesian coordinate system is rotated 66.2° clockwise around the star at 0,0 so that the northern semi-major axis becomes the positive x' axis. The y coordinates are then deprojected to y' by dividing by cos (66.0°) where 66° is the belt inclination to the line of sight. The conversion to au assumes a heliocentric distance of 7.704 pc. The 2023 and 2024 observations are separated by 357.7 days.

| DRM | Δ RA (mas) | Δ Dec (mas) | Sep (mas) | PA (°) | 2023 x' (au) | 2023 y' (au) | Sep (au) |
|---|---|---|---|---|---|---|---|
| 1 | -9599 | 8426 | 12773 | 311.28 | 89.24 | 101.95 | 135.49 |
| 2 | -9603 | 8419 | 12771 | 311.24 | 89.20 | 102.07 | 135.56 |
| 3 | -9585 | 8399 | 12744 | 311.23 | 89.46 | 102.23 | 135.30 |
| 4 | -9565 | 8414 | 12739 | 311.34 | 89.55 | 101.95 | 135.70 |
| **Mean** | -9588±17 | 8415±11 | 12757±18 | 311.27±0.05 | 89.12±0.12 | 101.85±0.27 | 135.33±0.26 |

**Table S3: Same as table S3, but for the 2024 candidate source**

| DRM | RA (mas) | Dec (mas) | Sep (mas) | PA (°) | 2024 x' (au) | 2024 y' (au) | Sep (au) |
|---|---|---|---|---|---|---|---|
| 1 | -9604 | 8538 | 12850 | 311.64 | 90.041 | 101.18 | 135.44 |
| 2 | -9623 | 8531 | 12860 | 311.56 | 90.051 | 101.56 | 135.73 |
| 3 | -9598 | 8507 | 12825 | 311.55 | 89.804 | 101.31 | 135.38 |
| 4 | -9578 | 8524 | 12821 | 311.67 | 89.862 | 100.84 | 135.07 |
| Mean | -9600±19 | 8525±13 | 12839±19 | 311.60±0.06 | 89.94±0.13 | 101.22±0.30 | 135.41±0.27 |

**Table S4**: **Best-fitting Keplerian orbit model for Fom cs2.** The 95% posterior reports the median value of the posterior and the uncertainties report the 95% credible interval (CI). The maximum posterior probability fit is also reported.

| Orbital Element | Posterior 95% CI | Best Fit |
|---|---|---|
| $a$ (au) | $144^{+364}_{-65}$ | 156 |
| $e$ | $0.66^{+0.25}_{-0.50}$ | 0.76 |
| $i$ (°) | $40^{+17}_{-19}$ | 21 |
| $\omega$ (°) | $182^{+93}_{-131}$ | 54 |
| $\Omega$ (°) | $179^{+148}_{-62}$ | 140 |
| $\tau$ | $0.91^{+0.08}_{-0.24}$ | 0.94 |
| Parallax (mas) | $129.82^{+0.96}_{-0.94}$ | 129.41 |
| $M_{tot}$ ($M_\odot$) | $1.92^{+0.04}_{-0.04}$ | 1.89 |

**Table S5: Fomalhaut cs1 astrometry in the sky plane.** All units are milliarcseconds.

| Ref. | 2004 RA | 2004 Dec | 2006 RA | 2006 Dec | 2010 RA | 2010 Dec | 2012 RA | 2012 Dec | 2013 RA | 2013 Dec |
|---|---|---|---|---|---|---|---|---|---|---|
| (*52*) | 8598±20 | 9209±14 | 8615±16 | 9352±16 | | | | | | |
| (*53*) | 8590±20 | 9190±20 | 8640±20 | 9360±20 | 8810±70 | 9790±70 | | | | |
| (*9*) | 8587±24 | 9175±17 | 8597±22 | 9365±19 | 8828±40 | 9822±44 | 8915±35 | 10016±37 | | |
| (*10*) | 8580±11 | 9198±11 | 8683±21 | 9341±21 | 8850±16 | 9824±16 | 8915±19 | 10024±20 | 9018±27 | 10173±25 |
| This work | | | | | | | | | 9091±36 | 10209±39 |
| **Median** | **8589±20** | **9194±16** | **8628±21** | **9356±20** | **8828±42** | **9822±44** | **8915±27** | **10020±29** | **9055±32** | **10191±32** |

**Table S6: Fomalhaut cs1 astrometry in the belt plane.** The belt in the x,y Cartesian coordinate system of the sky plane is rotated 66.2° clockwise around the star at 0,0 so that the northern semi-major axis becomes the positive x' axis. The y coordinates are then deprojected to y' by dividing by cos(66.0°) where 66° is the belt inclination to the line of sight. The conversion to au assumes a heliocentric distance of 7.704 pc. The Julian Date (JD) is the midpoint of observations (which may be spread over 2–3 days) for the UT month in the first column.

| UT Date yr–mo | JD - 2450000 | x' (au) | y' (au) | Δau from prior date | tangential (au) | radial (au) | Δt from prior date (sec) | tangential (km s$^{-1}$) | radial (km s$^{-1}$) |
|---|---|---|---|---|---|---|---|---|---|
| 2004–10 | 3304.25 | 91.484±0.134 | 78.623±0.157 | | | | | | |
| 2006–07 | 3935.36 | 92.761±0.135 | 77.756±0.174 | 1.544 | 1.490 | 0.404 | 5.45279×10$^7$ | 4.088 | 1.108 |
| 2010–09 | 5452.94 | 96.576±0.367 | 78.091±0.519 | 3.830 | 2.194 | 3.139 | 1.31119×10$^8$ | 2.503 | 3.581 |
| 2012-05 | 6078.17 | 98.317±0.238 | 77.941±0.328 | 1.747 | 1.259 | 1.211 | 5.40199×10$^7$ | 3.487 | 3.354 |
| 2013-05 | 6443.81 | 100.22±0.244 | 79.511±0.342 | 2.467 | 0.048 | 2.467 | 3.15913×10$^7$ | 0.189 | 11.682 |